\documentclass[12pt]{article}

\pdfoutput=1
\usepackage{color}
\usepackage{epsfig, palatino}
\usepackage{pstricks,pst-node,pst-tree}
\usepackage{epic}
\usepackage{mathrsfs}
\usepackage{ae} 
\usepackage[T1]{fontenc}
\usepackage[ansinew]{inputenc}
\usepackage{amsmath}
\usepackage{bbm}
\usepackage{amssymb}
\usepackage{graphicx}
\usepackage{ulem}

\definecolor{darkblue}{cmyk}{0.9,0.9,0,0}
\definecolor{darkgreen}{cmyk}{0.9,0,0.9,0}
\definecolor{blueblue}{cmyk}{0.73,0.28,0,0.5}
\definecolor{lightblue}{RGB}{55,171,200}
\definecolor{grey}{gray}{0.55}
\definecolor{pink}{cmyk}{0., 0.9859943977591037, 0.3571428571428571, 0.16000000000000003}
\definecolor{lightpink}{cmyk}{0., 0.5, 0.5, 0.}
\definecolor{lightgreen}{cmyk}{0.24175824175824182, 0., 0.9615384615384616, 0.28627450980392155}

\usepackage[colorlinks=true,linkcolor=darkblue,citecolor=darkblue,urlcolor=darkblue]{hyperref}
\usepackage{cite}
\usepackage{hyperref}
\usepackage{wasysym}
\usepackage{varioref}
\usepackage{makeidx}
\usepackage[english]{babel}
\usepackage{simplewick}
\usepackage{array}
\usepackage{multirow}
\usepackage{tabularx}
\usepackage[font={small}]{caption}
\usepackage[font={small}]{subcaption}

\def\({\left(}
\def\){\right)}
\def\[{\left[}
\def\]{\right]}
\def\<{\langle}
\def\>{\rangle}

\newcommand{\TTB}{\textrm{T}\bar{\textrm{T}}}

\newcommand{\cO}{\mathcal O}

\newcommand{\sn}{\text{sn}}

\newcommand{\Tl}{T_l}

\newcommand{\la}[1]{\label{#1}} 

\newcommand{\beq}{\begin{equation}}
\newcommand{\eeq}{\end{equation}}
\newcommand{\beqq}{\begin{equation*}}
\newcommand{\eeqq}{\end{equation*}}
\newcommand\beqa{\begin{eqnarray}}
\newcommand\eeqa{\end{eqnarray}}

\newcommand{\nn}{\nonumber}

\newcommand{\ii}{i}

\makeindex

\begin{document}

\thispagestyle{empty}

\renewcommand{\thefootnote}{\fnsymbol{footnote}}
\setcounter{page}{1}
\setcounter{footnote}{0}
\setcounter{figure}{0}
\begin{center}
$$$$
{\Large\textbf{\mathversion{bold}
Thermodynamic Bethe Ansatz past turning points:\\the (elliptic) sinh-Gordon model}\par}

\vspace{1.0cm}

\textrm{Luc\'ia C\'ordova$^\text{\tiny 1}$, Stefano Negro$^\text{\tiny 2}$, Fidel I. Schaposnik Massolo$^\text{\tiny 3}$}
\\ \vspace{1.2cm}
\footnotesize{\textit{
$^\text{\tiny 1}$ Institut de Physique Th\'eorique Philippe Meyer, Laboratoire de Physique de l'\'Ecole Normale Sup\'erieure\\ PSL University, CNRS, Sorbonne Universit\'es, UPMC Univ. Paris 06\\ 24 rue Lhomond, 75231 Paris Cedex 05, France.\\
$^\text{\tiny 2}$ Center for Cosmology and Particle Physics, New York University, New York, NY 10003, U.S.A.\\
$^\text{\tiny 3}$ Institut des Hautes \'Etudes Scientifiques, 35 route de Chartres, 91440 Bures-sur-Yvette, France.
}  
\vspace{4mm}
}

\par\vspace{1.5cm}

\textbf{Abstract}\vspace{2mm}

\end{center}

\noindent

We analyze the Thermodynamic Bethe Ansatz (TBA) for various integrable S-matrices in the context of generalized $\TTB$ deformations. We focus on the sinh-Gordon model and its elliptic deformation in both its fermionic and bosonic realizations.
We confirm that the determining factor for a turning point in the TBA, interpreted as a finite Hagedorn temperature, is the difference between the number of bound states and resonances in the theory. 
Implementing the numerical pseudo-arclength continuation method, we are able to follow the solutions to the TBA equations past the turning point all the way to the ultraviolet regime. We find that for any number $k$ of resonances the pair of complex conjugate solutions below the turning point is such that the effective central charge is minimized. As $k\to\infty$ the UV effective central charge goes to zero as in the elliptic sinh-Gordon model. Finally we uncover a new family of UV complete integrable theories defined by the bosonic counterparts of the $S$-matrices describing the $\Phi_{1,3}$ integrable deformation of non-unitary minimal models $\mathcal M_{2,2n+3}$.

\setcounter{page}{1}
\renewcommand{\thefootnote}{\arabic{footnote}}
\setcounter{footnote}{0}

 \def\nref#1{{(\ref{#1})}}

\newpage

\tableofcontents

\parskip 5pt plus 1pt   \jot = 1.5ex

\newpage

\section{Introduction}

Quantum Field Theory (QFT) provides a framework capable of describing a wealth of different phenomena with astonishing accuracy. As such, it constitutes one of the pillars sustaining our modern understanding of Nature. It is hard to overstate the importance of obtaining a complete and satisfactory understanding of QFTs. These can be thought as populating a vast landscape, the \textit{theory space}, which can be explored with the help of methods such as the Renormalization Group (RG). Conformal Field Theories (CFTs) play a pre-eminent role in this description as fixed points\footnote{More precisely, the fixed points of the RG flows are scale invariant theories. For a moderately recent account on the distinction between scale and conformally invariant theories, see \cite{Nakayama:2013is}.} of the RG flows. From these, one can then explore the theory space by perturbing the corresponding CFT with a given set of operators. Depending on their scaling dimension, the deforming operators and the associated RG flows are arranged in three distinctive classes: \textit{relevant}, \textit{marginal} and \textit{irrelevant}. While considerable work has been devoted to the first two, the irrelevant deformations have received much less attention and to this date their high-energy physics constitute a largely unexplored territory. There is a very good reason for this. General arguments of renormalization show that the perturbative analysis of irrelevant deformations leads to an accumulation of UV divergences which necessarily require an infinite number of counter-terms, destroying the predictive power of the theory. Ultimately the high-energy behavior of such theories is completely ambiguous and therefore confined to the realm of effective field theory. 

The grim destiny looming over the UV behavior of irrelevant deformations is not inescapable, at least not for all theories. In the last few years, certain special classes of irrelevant deformations in two space-time dimensions were shown to be under exceptionally good control, even deep in the UV. The poster child of these is the so-called $\TTB$ deformation \cite{Smirnov:2016lqw,Cavaglia:2016oda}, triggered by the composite irrelevant operator built from the components of the energy-momentum tensor \cite{Zamolodchikov:2004ce}. This deformation possesses several remarkable properties. It was shown to be intimately related to two-dimensional gravity \cite{Dubovsky:2017cnj,Dubovsky:2018bmo,Conti:2018tca,Ishii:2019uwk}, random geometry \cite{Cardy:2018sdv}, string theory \cite{Cavaglia:2016oda,Baggio:2018gct,Hashimoto:2019wct,Sfondrini:2019smd,Callebaut:2019omt,Tolley:2019nmm} and holography \cite{McGough:2016lol,Giveon:2017nie,Chakraborty:2019mdf}. The $\TTB$ deformed theories are solvable, in the sense that physical observables of interest, such as the finite-volume spectrum, the $S$-matrix \cite{Smirnov:2016lqw,Cavaglia:2016oda,Dubovsky:2017cnj} and the partition functions \cite{Cardy:2018sdv,Dubovsky:2018bmo}, can be determined exactly in terms of the corresponding undeformed quantities. This high degree of control means that it is possible to follow the irrelevant flow stemming from the undeformed CFT all the way to the UV, effectively reversing the renormalisation group trajectory, and to obtain exact results on the UV physics. The latter are remarkable: the finite-size density of states grows exponentially at high energies, in a Hagedorn fashion \cite{Hagedorn:1965st} reminiscent of string theories \cite{Dubovsky:2012sh,Dubovsky:2012wk,Caselle:2013dra}. This exceptional feature is already manifest in the short-scale behavior of the finite-size groundstate energy. If we let $R$ denote the circumference of the circle on which the space component of two-dimensional space-time is compactified, then the groundstate energy $E_{\alpha}(R)$ of a $\TTB$ deformed system obeys the functional equation \cite{Smirnov:2016lqw,Cavaglia:2016oda}
\beq
    E_{\alpha}(R) = E_{\alpha=0}(R - \alpha E_{\alpha}(R))\;,
    \label{eq:flow_eq_TTbar}
\eeq
where $\alpha$ is the deformation parameter. Depending on the sign of $\alpha$, the function $E_{\alpha}(R)$ determined by \eqref{eq:flow_eq_TTbar} either presents a square-root singularity at a positive radius $R_\ast \sim \vert\alpha\vert^{-1/2}$ (for $\alpha < 0$) or possesses no short-scale singularity at all (for $\alpha > 0$). Neither of these is compatible with Wilson's paradigm of local QFTs \cite{Wilson:1973jj}, leading us to the conclusion that $\TTB$ deformed theories cannot be considered conventional UV-complete theories and that, thanks to their robust features, they represent a sensible extension of the Wilsonian notion of a local QFT.

The compelling features of $\TTB$ deformations are in large part shared by larger families of two-dimensional theories that we will refer to collectively as \textit{solvable irrelevant deformations}. Important examples are the Lorentz-breaking $\textrm{J}\bar{\textrm{T}}$ deformations \cite{Guica:2017lia}, triggered by the composite operator $\textrm{J}\bar{\textrm{T}}$ that can be defined whenever the undeformed theory possesses a conserved holomorphic $U(1)$ current $J_{\mu}$. Another very large and important family of solvable irrelevant deformations is available if one considers integrable QFTs \cite{Smirnov:2016lqw}. In this case we have access to an infinite space of conserved currents $T_{\mu\nu}^{(s)}$ labeled by a half-integer spin\footnote{To be more precise, in complex coordinates $(z,\bar{z})$ the components $T_{zz}^{(s)}$ and $T_{\bar{z}\bar{z}}^{(s)}$ have spin $s+1$, while $T_{z\bar{z}}^{(s)}$ and $T_{\bar{z}z}^{(s)}$ have spin $s-1$. The requirement of locality restricts the spins to be either integers or half-integers. The specific subset $\mathbb S$ in which $s$ takes values is a characteristic of the integrable QFT under consideration.} $s\in \mathbb S\subset \mathbb{Z}/2$, and each of these can be used to construct a two-parameter family of bi-linear irrelevant operators $\TTB^{(s,s')} \equiv \varepsilon^{\mu\rho} \varepsilon^{\nu\sigma} T_{\mu\nu}^{(s)} T_{\rho\sigma}^{(s')}$ (here $\varepsilon^{\mu\nu}$ is the totally antisymmetric Levi-Civita symbol in two dimensions). Integrable QFTs deformed by these operators were shown to preserve their integrability and the corresponding set of local conserved charges \cite{Smirnov:2016lqw,Conti:2018jho,LeFloch:2019wlf}, allowing the use of powerful exact, non-perturbative methods in their analysis. Hence, this vast family of theories represents the perfect playground on which to study the ultraviolet properties of theories not complying with the Wilsonian QFT paradigm. Of primary importance is the question concerning the nature of the short-scale physics of these models, with particular focus on the status of their local structures and the mechanisms responsible for the appearance of a Hagedorn temperature. Understanding the physics underlying these matters remains the most important open problem in this context, as well as one of the main motivations for the present work.

Amongst the vast set of irrelevant deformations triggered by the operators $\TTB^{(s,s')}$, a sub-family of Lorentz-breaking ones obtained by fixing $s'=1$ was introduced and studied in \cite{Conti:2019dxg}, while \cite{Hernandez-Chifflet:2019sua} and \cite{Camilo:2021gro} initiated the analysis of integrable QFTs deformed by linear combinations of the operators $\TTB^{(s)}\equiv \TTB^{(s,s)}$. In this work we will focus our attention on particular instances of the latter type of deformations that we will refer to as \textit{generalized $\TTB$ deformations}. The most convenient setting to analyze these is provided by the factorised scattering framework \cite{Zamolodchikov:1989hfa}. As explained thoroughly in \cite{Camilo:2021gro}, given an integrable QFT with elastic two-body $S$-matrix $S_0(\theta)$, the space of its generalized $\TTB$ deformations stands in one-to-one correspondence with the space of \textit{CDD factor} deformations of its $S$-matrix,
\beq
    S(\theta) = S_{0}(\theta)\Phi(\theta)\;.
    \label{eq:CDD_def_S_mat}
\eeq
Here the CDD factor \cite{Castillejo:1955ed} $\Phi(\theta)$ is a scalar phase factor, arbitrary up to the request that its presence does not violate the unitarity, crossing symmetry, analyticity and macro-causality conditions of the $S$-matrix \cite{Eden:1966dnq,Iagolnitzer:1974wz,Iagolnitzer:1994xv}. Ultimately these requirements force the CDD factor to be a meromorphic function of the rapidity $\theta$ of the form
\beq
    \Phi(\theta) = \Phi_{\textrm{prod}}(\theta)\Phi_{\textrm{entire}}(\theta)\; ,
\eeq
where the first factor incorporates all the poles at finite $\theta$, whose number $N$ is arbitrary (and possibly infinite)
\beq
    \Phi_{\textrm{prod}}(\theta) = \prod_{j=1}^N \frac{\sinh\theta + \sinh\theta_j}{\sinh\theta - \sinh\theta_j}\;.
    \label{eq:phipole}
\eeq
The second factor represents an entire function of $\theta$ of the form
\beq
    \Phi_{\textrm{entire}}(\theta) = \exp\left[-\ii\,\sum_{s\in \mathbb S} a_s\,\sinh(s\theta)\right]\;,
    \label{eq:phientire}
\eeq
with $\mathbb S$ being the characteristic set of spins mentioned above. The series in the exponent of \eqref{eq:phientire} is assumed to be convergent for any $\theta$. Finally, the condition of macro-causality \cite{Iagolnitzer:1994xv} restricts possible positions of the poles in \eqref{eq:phipole} to either the imaginary axis $\textrm{Re}\, \theta_j = 0$, or to the strips $\textrm{Im}\, \theta_j \in \left[-\pi,0\right] \mod\, 2\pi$. Each pole has a standard interpretation in terms of the spectrum of the theory. The purely imaginary ones signal, when their imaginary part is positive, the presence of bound states of mass $2m\cosh\left(\theta_j/2\right)$, where $m$ stands for the mass of the lightest excitation in the theory. The poles $\theta_j$ with $\textrm{Re}\,\theta_j\neq 0$ and $-\pi\leq\textrm{Im}\,\theta_j<0$ lie in the so-called \textit{unphysical strip}\footnote{This corresponds to the region of the complex center-of-mass energy s-plane reached by analytically continuing the scattering amplitude through the two-particle branch cut.} and are associated with unstable particles, also known as \textit{resonances}. The real and imaginary parts of their complex mass $2m\cosh\left(\theta_j/2\right)$ are identified, respectively, with the center-of-mass energy and the width of the resonance. The last possibility is to have poles lying on the negative imaginary axis. These are usually referred to as \textit{virtual states} \cite{perelomov1998quantum}, and while they do not have a clear interpretation in terms of particles their presence effects an increment of the scattering phase as a function of $\theta$ at low energies.

In this work, we continue the program laid down in \cite{Camilo:2021gro} by studying a number of models obtained by deforming a trivial $S$-matrix $S_0(\theta) = \pm 1$ with $\Phi_\text{entire}=1$ and a
product CDD factor \eqref{eq:phipole},
\beq
    S(\theta) = \pm\prod_{j=1}^N \frac{\sinh\theta + \sinh\theta_j}{\sinh\theta - \sinh\theta_j}\;,
\eeq
where the poles $\big\{\theta_j\big\}_{j=1}^{N}$ all lie in the unphysical strip, thus corresponding to resonances. Here the upper (lower) sign corresponds to the fermionic (bosonic) TBA statistics (see \S2 of \cite{Zamolodchikov:1990bk} for an explanation of this point).
We will be especially interested in a particular exponent of this class with $N=\infty$ resonances: the \textit{elliptic sinh-Gordon model} \cite{Mussardo:1999ee}. 

Our interest in this particular model is two-fold. On one hand we can view it as the effect of deforming a free theory with a particular combination of generalized $\TTB$ operators as discussed above. On the other hand its periodic nature allows us to make direct contact with the periodic $S$-matrices arising in the recently revived $S$-matrix bootstrap program \cite{Paulos:2016fap,Paulos:2016but}, and more generally to theories with a large number of resonances.

In the $S$-matrix bootstrap program one explores the space of consistent $S$-matrices by imposing general principles like unitarity, crossing symmetry and analyticity. When restricting to two spacetime dimensions, integrable theories naturally appear at the boundary of the allowed space of theories\cite{Paulos:2016fap,Paulos:2016but,He:2018uxa,Cordova:2018uop,Paulos:2018fym,EliasMiro:2019kyf,Homrich:2019cbt,Cordova:2019lot,Bercini:2019vme}. In this way one can rediscover known integrable models, but also obtain integrable $S$-matrices with no Lagrangian realization. A paradigmatic example of the second situation is the $O(N)$ \textit{periodic Yang-Baxter model}. Its integrable $S$-matrix was proposed in \cite{Hortacsu:1979pu} and later found sitting at a special point of the \textit{monolith} parametrizing the space of theories with a global $O(N)$ symmetry and no bound states \cite{Cordova:2019lot}. This model, as the elliptic sinh-Gordon one, has infinite resonances spaced periodically in the rapidity plane (hence its name). The global symmetry makes it however more complicated than the single-particle models analyzed in this article, so we restrict our attention to the simplest periodic $S$-matrices and leave the study of global symmetry analogues for future work. 

The plan of this paper is as follows. In section~\ref{sec:elliptic_model} we review the elliptic sinh-Gordon models and the Thermodynamic Bethe Ansatz (TBA), the primary tool we will use to study them. Section~\ref{sec:numerical_methods} contains a review of the numerical approaches to the solution of the TBA equation, with a special emphasis on the application of the pseudo-arclength continuation method required to overcome the limitations of the more traditional iterative methods typically used in the literature. Most of our new results are presented in section~\ref{sec:results}. After considering the standard sinh-Gordon model as a warm-up we tackle its elliptic deformation, in both its fermionic and bosonic flavors, devoting special attention to the UV and IR behavior of the effective central charge. Section~\ref{sec:minimal_models} initiates the analysis of CDD deformations of theories with more than a single stable particle by looking at the simple examples provided by the $S$-matrices of the $\Phi_{1,3}$ integrable deformation of the non-unitary minimal models and their bosonic counterparts. We conclude in section~\ref{sec:discussion} with a discussion of our results and an outlook for future research directions. Three appendices provide some background on elliptic functions, an interesting albeit somewhat mysterious relation between the TBA kernel and the convolution term in the elliptic models, and a more detailed description of our implementation of the pseudo-arclength continuation method.

\section{Elliptic sinh-Gordon models and the Thermodynamic Bethe Ansatz}\label{sec:elliptic_model}

\begin{figure}
\centering
\includegraphics[width=.45\textwidth]{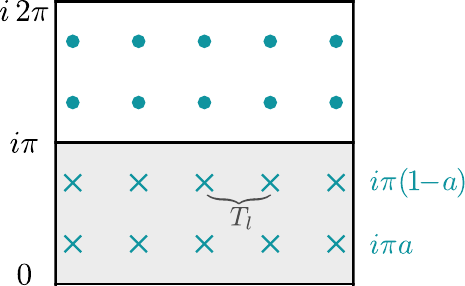}\vspace{0.3cm}
\caption{Analytic structure of the elliptic sinh-Gordon $S$-matrix in the complex rapidity plane. Simple zeros (poles) are depicted by crosses (dots). The shaded region is the physical rapidity strip $0\leq \text{Im}(\theta) \leq\pi$ which contains two crossing symmetric towers of infinite resonances separated in the real direction by the period $T_l$. 
}
\label{fig_S_plane}
\end{figure}

The main model of interest in the present work is an elliptic deformation of the sinh-Gordon model. Its $S$-matrix was proposed in \cite{Mussardo:1999ee} and reads
\beq
S^\text{f,b}_{a,l}(\theta)=\pm \frac{\sn_l \(2 \ii K_l \,  \theta/\pi\) + \sn_l(2K_l \, a)}{\sn_l \(2 \ii K_l \, \theta/\pi\) - \sn_l(2K_l \, a)}\,, \label{S_ellipt}
\eeq
where $a$ is the coupling constant, the modulus $l$ parametrizes the periodicity in real rapidity, $\sn$ is the Jacobi sine function and $K_l$ is the complete elliptic integral 
\beq
    K_l=\int_0^{\pi/2} \frac{d\phi}{\sqrt{1-l^2 \sin^2\phi}}\;.
\eeq
The upper (lower) sign in \eqref{S_ellipt} corresponds to the fermionic (bosonic) case. This $S$-matrix has two periods, the usual one along the imaginary direction arising from crossing symmetry and unitarity $S(\theta)=S(\theta+2i\pi)$ and an unusual one in the real direction $S(\theta)=S(\theta+\Tl)$. The latter period is given in terms of elliptic integrals,
\beq
\Tl=\pi\frac{K_{\sqrt{1-l^2}}}{K_l}\,. \label{period}
\eeq
The elliptic sinh-Gordon model contains an infinite tower of resonances spaced periodically in the rapidity plane. We see them as complex zeros\footnote{Due to the unitarity condition $S(\theta)S(-\theta) = 1$, complex zeroes in the physical strip are in one-to-one correspondence with poles in the unphysical strip. Thus their existence signals the presence of a resonance in the spectrum of the theory.} in the physical rapidity strip defined by $\text{Im}(\theta) \in (0,\pi)$. The location of the ``fundamental'' zeros is $\theta_n=i a \pi +n \Tl$ with $n\in \mathbb Z$ and the arrangement of the remaining zeros and poles outside the physical strip follows from the crossing and unitarity conditions of the $S$-matrix.  Figure~\ref{fig_S_plane} shows the analytic structure in the physical strip.

In order to ensure we have only zeros inside the physical strip --and, consequently, that the model possesses no bound-state-- we need to limit the coupling constant $a$ to take values in the interval $[0,1]$. Moreover, since the model is invariant under $a\leftrightarrow1-a$, we will consider $a\in[0,1/2]$ without any loss of generality. The limit $a\rightarrow0$ leads to free theory, $S^\text{ f,b}_{0,l}=\mp1$. At zero modulus $l\rightarrow0$ --equivalent to infinite period $\Tl\rightarrow\infty$--  we recover the usual sinh-Gordon $S$-matrix with a single resonance
\beq
S^\text{ f,b}_{a,0}(\theta) = S^\text{f,b}_\text{shG} = \pm\frac{\sinh\theta-i\sin\(\pi a\)}{\sinh\theta+i\sin\(\pi a\)}\,.
\eeq

Even though the $S$-matrix \eqref{S_ellipt} approaches well-known theories in the above limits, its physical nature is still mysterious, \textit{e.g.} there is no known Lagrangian formulation for the theory. To gain more information about these models and, in particular, about their ultraviolet behaviour, we will use the Thermodynamic Bethe Ansatz (TBA) to compute the groundstate energy and the effective central charge of the theory compactified in a circle of circumference $R$. In the following we give a brief review of the method, for a more comprehensive treatment the reader is referred to \cite{Zamolodchikov:1989cf,vanTongeren:2016hhc} and references therein.

For the simplest scenario where we have only one type of stable particle of mass $m$, the TBA equation is
\beq\label{tba_equation}
\epsilon(\theta) = r \cosh\theta \mp \int\limits_{-\infty}^\infty \varphi(\theta-\theta') \ln\[1 \pm e^{-\epsilon(\theta')}\] \frac{d\theta'}{2\pi}\,,
\eeq
where $r=m R$, the $S$-matrix enters through the kernel $\varphi(\theta)=-i\,\partial_\theta \ln S(\theta)$ and one solves for the pseudo-energy $\epsilon(\theta)$. The upper and lower signs refer to the fermionic and bosonic cases, respectively, and in what follows we will sometimes refer for brevity to the $L$--functions $L(\theta) = \pm \ln\left[1\pm e^{-\epsilon(\theta)}\right]$.

Except for very rare cases such as the standard $\TTB$ deformation, one needs to solve the TBA equations \eqref{tba_equation} numerically. As we will see, the $S$-matrices studied in this paper require a refined numerical method which we review in section~\ref{sec:numerical_methods}. In any case, once a solution $\epsilon(\theta)$ to \eqref{tba_equation} has been found, the effective central charge and groundstate energy are computed from the expressions
\beq\label{central_charge}
\tilde c^\text{\,f,b}(r)= \pm\frac{6}{\pi}\int\limits_{-\infty}^{\infty} r \cosh(\theta)\ln\[1\pm e^{-\epsilon(\theta)}\] \frac{d\theta}{2\pi}\,,\qquad E_0(R)=-\frac{\pi}{6}\frac{\tilde c(r)}{R}\,.
\eeq
Due to the absence of massless excitations, the infrared limit $r\to \infty$ consists of a trivial theory. Accordingly, the effective central charge tends to zero in a precise and universal fashion, controlled by the modified Bessel function $\text{K}_1$,
\beq
\tilde c(r) \underset{r\to\infty}{\sim}\frac{6\,r}{\pi^2}\, \text{K}_1(r) \underset{r\to\infty}{\sim} 3\sqrt{\frac{2r}{\pi^2}} e^{-r} \;.
\eeq

In the usual situation in which the theory under consideration is UV complete, one can relate the central charge $c_\text{UV}$ of the CFT controlling the UV behavior to the effective central charge in the $r\to 0$ limit 
\beq
\lim_{r\rightarrow0}\, \tilde c(r)=c_\text{UV}-12(\Delta_\text{min}+\bar\Delta_\text{min})\,,
\eeq
where $\Delta_\text{min}$ is the lowest dimension in the UV theory. The exact value for $\tilde c(r=0)$ is model-dependent but can be obtained fairly easily with the help of the so-called \textit{dilogarithm trick}\footnote{To the best of our knowledge, this first appeared in \cite{tsvelick1983exact} where it was employed to derive the heat capacity of the Anderson model. For an application of this trick to the TBA of relativistic theories, see \cite{Zamolodchikov:1989cf}.} \cite{Klassen:1989ui}. The key point in this calculation is the observation that as $r\rightarrow0$ the pseudo-energy becomes constant in a region $|\theta|<\ln(2/r)$. One therefore proceeds to solve the TBA equation for $r=0$ and a constant pseudo-energy $\epsilon$,
\beq\label{r_to_0_tba}
\epsilon=\mp\,\Omega \ln\(1\pm e^{-\epsilon}\) \qquad\text{with}\qquad \Omega = \int\limits_{-\infty}^\infty \varphi(\theta)\frac{d\theta}{2\pi}\,.
\eeq
The UV effective central charge is finally computed in terms of Rogers' dilogarithm function $\text{L}(x)=\text{Li}_2(x)+\tfrac{1}{2}\ln(x)\ln(1-x)$, as
\beq
\tilde c(0)=\frac{6}{\pi^2}\,\times
\begin{cases}
\text{L}\(\dfrac{1}{1+e^{\epsilon}}\) & \text{fermionic}\\
\text{L}\(e^{-\epsilon}\) & \text{bosonic}
\end{cases}
\label{c_dilog}
\eeq

In the following section we will describe the different algorithms that can be used to numerically solve the TBA equation \eqref{tba_equation}. These methods, specifically the pseudo-arclength continuation, were implemented to extract the results that will be presented in section~\ref{sec:results} and section~\ref{sec:minimal_models}.

\section{Numerical methods}
\label{sec:numerical_methods}

We would like to solve the TBA equation \eqref{tba_equation} numerically for the elliptic sinh-Gordon models with kernel $\varphi_{a,l}(\theta)$ given by %
\beq\la{kernel_ellshG}
\varphi_{a,l}(\theta) = -i\,\partial_\theta \ln S_{a,l}(\theta) = \frac{4 K_l}{\pi} \frac{{\rm cn}_l(2i K_l\,\theta/\pi) {\rm dn}_l(2i K_l\, \theta/\pi) {\rm sn}_l(2a K_l)}{{\rm sn}_l(2a K_l)^2 - {\rm sn}_l(2i K_l\, \theta/\pi)^2}\,,
\eeq
where ${\rm sn}_l$, ${\rm cn}_l$ and ${\rm dn}_l$ are respectively the basic Jacobi elliptic sine, cosine and delta amplitude. Note the kernel is identical for the fermionic and bosonic cases, the only difference between these resides in the TBA equation itself.

Already in \cite{Zamolodchikov:1989cf} it was noted that this equation can be solved numerically by an iterative process, whereby we start with an initial approximation $\epsilon^{(0)}(\theta)$ for the pseudo-energy, and produce increasingly more accurate ones through
\beq\label{iterative_update}
\epsilon^{(k+1)}(\theta) = r \cosh\theta \mp \int\limits_{-\infty}^\infty \varphi_{a,l}(\theta-\theta') \ln\left(1 \pm e^{-\epsilon^{(k)}(\theta')}\right) \frac{d\theta'}{2\pi} \qquad\text{for}\qquad k = 0, 1, \dots\,.
\eeq
This process is not guaranteed to converge, but if it does the iterations can be stopped once the desired level of accuracy is reached for the solution.

The only peculiarity in the numerical formulation of our problem is that the kernel $\varphi_{a,l}(\theta)$ has a real period, and therefore the result of the convolution is also periodic with period $\Tl$. The pseudo-energy being an even function of $\theta$, for numerical purposes it is thus convenient to discretize an interval in the real-$\theta$ line centered at $\theta = 0$, whose size is an even number of $\Tl$ periods. More concretely, we take the following discretization of the interval $[-M \Tl, M \Tl)$,
\beq\label{theta_discretization}
\theta_i = -M \Tl  + \frac{i-1}{N}\Tl \qquad\text{for}\qquad i = 1, 2, \dots, 2NM \qquad\text{and}\qquad N, M \in \mathbb{N} \,,
\eeq
which then allows us to immediately apply the circular convolution theorem to improve the efficiency of the computation of convolutions. The resulting algorithm is then $\cO(N \log N + N M)$, since the number of iterations required to converge to a solution is typically $\cO(1)$.

The procedure described above is well-suited to study families of solutions depending on continuous parameters. Indeed, one may use a solution found at a given point in parameter space as an initial approximation to start the iterative process at another nearby point. For our purposes, this is true both for the coupling constant $a$ and the radius $r$, since the discretization \eqref{theta_discretization} does not depend on their values. We will however concentrate mostly on the latter, \textit{i.e.} we will consider families of solutions related by a continuous variation of the radius $r$.

For large values of $r$ we can neglect the convolution term in the TBA equation and start from the initial approximation $\epsilon^{(0)}(\theta) = r \cosh\theta$. We may then slowly decrease the radius, but in some cases it turns out that the iterative process takes longer to converge, and eventually stops converging altogether, for small enough values of $r$. However, failure to converge is only indicative of a numerical instability and should not be confused with an indication that solutions to \eqref{tba_equation} do not exist beyond what can be found by iteration. Indeed, this numerical instability could in principle be the result of the family of solutions approaching a singular point, but this is not necessarily always the case as will be the case \textit{e.g.} in section~\ref{sec:minimal_models}.

To address the issue above, we follow\footnote{Our implementation in Mathematica 12.3 is a simplified version of the algorithm described there, relying at its core on the {\tt FindRoot} function to numerically solve the resulting systems of coupled nonlinear equations.} \cite{Camilo:2021gro} and implement a pseudo-arclength continuation method \cite{allgower2012numerical, allgower1997numerical} on the parameter $r$. In other words, keeping the discretization \eqref{theta_discretization} we directly write \eqref{tba_equation} as a set of $2 N M$ coupled non-linear equations on as many variables, $\epsilon_i = \epsilon(\theta_i)$ for $i = 1, 2, \dots 2 N M$. We then introduce an arclength parameter $s$ such that a family of solutions is described by a curve $c(s) = (\epsilon_i(s), r(s))$, and impose the normalization condition $||\tfrac{dc}{ds}|| = 1$. The whole process can be viewed as a numerical stabilization mechanism for the iterative solution of \eqref{tba_equation}, allowing us to follow the solution curve past turning points and bifurcations which may constitute the sources of instability for the more traditional iterative method. We spell out further the pseudo-arclength continuation procedure in Appendix \ref{app:pseudo-arclength}, while the reader is referred to \cite{Camilo:2021gro} for more details and to \cite{allgower2012numerical, allgower1997numerical} and references therein for the theoretical background on this procedure.

~

\noindent\textbf{Complex pseudo-energies} While the TBA equation \eqref{tba_equation} involves a real-valued pseudo-energy $\epsilon: \mathbb{R} \mapsto \mathbb{R}$, we can at least in principle consider the natural generalization to complex-valued pseudo-energies $\epsilon: \mathbb{R} \mapsto \mathbb{C}$. From a purely mathematical point of view this is a well-posed problem, so it may therefore be interesting to explore the space of complex solutions to \eqref{tba_equation}, of which the real-valued physical solutions turn out to be a non-trivial subspace.

For this purpose, the numerical algorithms described above require only minimal modifications. Indeed, the iterative method remains unchanged, whereas in the pseudo-arclength continuation method we simply need to double the number of variables to account for both the real and imaginary parts of $\epsilon(\theta_i)$, updating normalization conditions accordingly.

~

\noindent\textbf{Bifurcation detection and branch-switching} When studying families of solutions to the TBA equation related by the variation of the radius $r$ it is important to identify bifurcation points, \textit{i.e.} points whose immediate vicinity, no matter how small, includes multiple branches of solutions. There is a considerable literature dedicated to this issue in the context of numerical continuation methods \cite{allgower2012numerical, allgower1997numerical}. In particular, when performing pseudo-arclength continuation we can detect bifurcation points by keeping track of the augmented Jacobian determinant, which in our case reads
\beq\label{augmented_jacobian}
J = \begin{vmatrix}
\frac{\partial F(\epsilon_i, r)}{\partial \epsilon_i} & \frac{\partial F(\epsilon_i, r)}{\partial r} \\
\frac{d\epsilon_i}{ds} & \frac{dr}{ds}
\end{vmatrix}\,,
\eeq
where $F(\epsilon_i, r) = 0$ is the discretized TBA equation. As we construct a given family of solutions, the quantity $J$ defined above changes signs whenever a bifurcation point is crossed. Therefore, as long as we take small enough steps while performing the continuation we can make sure no bifurcations have inadvertently been skipped. Moreover, once a bifurcation point has been detected we can pinpoint its precise location using binary search to increasingly tighten the bounds on the parameters where the sign-change occurs.

Finally, various techniques have been developed in order to ``switch branches'' for the purpose of exploring the intersecting family of solutions once a bifurcation point has been identified. In our case, to accomplish this it is sufficient to introduce small perturbations to the initial starting point provided to the underlying Newton-like solver. This allows us to exhaustively explore the space of complex solutions to \eqref{tba_equation} which are connected to the $r\to\infty$ asymptotic solution $\epsilon(\theta) = r \cosh\theta$.

\section{Models with resonances: turning points in the TBA}\label{sec:results}

\subsection{Real branches and asymptotic properties}\label{sec:asymptotic_analysis}

Before presenting the numerical results for the different models, we review some asymptotic properties of the TBA equation \eqref{tba_equation}. 

The large radius analysis of the fermionic TBA equation in \cite{Camilo:2021gro} shows that 
the pseudo-energy admits two possible behaviors for $r\gg 1$: the first one is the standard asymptotic $\epsilon(\theta)\sim r\cosh(\theta)$, whereas the second one is $\epsilon(\theta)\sim -r f(\theta)$, where $f(\theta)>0$ for $\theta\in\Theta$ and $\Theta$ is a finite subset of the real line.\footnote{A third possibility with $\epsilon(\theta)\sim r^{-1} \cosh(\theta)$ is only viable for $\TTB$ deformations, as opposed to generalized $\TTB$ deformations with general integrable $S$-matrices.} The latter case is only allowed when the kernel $\varphi(\theta)$ is positive on the real line and satisfies the inequality 
\beq
\Omega = \int\limits_{-\infty}^\infty \varphi(\theta)\frac{d\theta}{2\pi} > 1\,.
\la{L1_norm}
\eeq
The integral above counts the number of resonances minus the number of bound states, so having two more resonances than bound states in the theory will generally allow for a second branch of solutions. For example, a single CDD-zero $S$-matrix (\textit{i.e.} having one resonance and zero bound states), such as the sinh-Gordon model, has a single branch of solutions to the TBA equation interpolating between the expected UV and IR behaviours. On the other hand the two CDD-zeros models studied in \cite{Camilo:2021gro} have both real branches.\footnote{Note that this argument does not constrain the local properties of the solutions, only their $r\rightarrow\infty$ asymptotic behavior. One might, in theory, observe any number of branch points, of any order, at finite $r$, so long as only two of the branches flow to $r\rightarrow\infty$ (or one, in case \eqref{L1_norm} is violated). Numerical evidence coming from various models, however, shows that the simplest possible case of a single square-root branch point (or no branch point if \eqref{L1_norm} is false) is the one presenting itself.}

Now let us turn to the bosonic TBA equation. The asymptotic analysis follows from the fermionic one through an appropriate redefinition of pseudo-energies and kernel. The bosonic TBA equation with pseudo-energy $
\epsilon^+(\theta)$ and kernel $\varphi^+(\theta)$ can be recast into a fermionic TBA equation with pseudo-energy $\epsilon^-(\theta)$ and kernel $\varphi^-(\theta)$ provided we define
\beqa
\epsilon^-(\theta)&=&\ln\[e^{\epsilon^+(\theta)}-1\]\,,\nn\\
\varphi^-(\theta)&=&\varphi^+(\theta)+2\pi\delta(\theta)\,.\la{eq:map_bos}
\eeqa

The delta function in the kernel above effectively adds a resonance to the counting in \eqref{L1_norm}. This interpretation of the bosonic TBA as a fermionic one with an added resonance is supported by the identity 
\beq
    \lim_{u\to 0^-}\log\left[\frac{\ii\sin u + \sinh\theta}{\ii \sin u - \sinh\theta}\right] = \ii\pi\,\textrm{sign}(\theta)\;,
    \label{eq:massless_resonance}
\eeq
showing that the $\delta$-function in \eqref{eq:map_bos} arises as a resonance CDD factor in the limit of the resonance mass going to zero. Consequently, we expect the single-particle bosonic TBA to allow for a second branch of solutions already in the presence of a single CDD-zero in the $S$-matrix. The leading asymptotic behaviour in the second branch for various quantities of interest is summarized in table~\ref{tab_asympt}.

\begin{table}
\centering
\begin{tabular}{ |c||c|c|  }
 \hline
 & \textbf{Fermionic}   & \textbf{Bosonic}\\
 \hline
  \hline
 $\epsilon(\theta)\sim$ & $-r f(\theta)$ & $e^{-r f(\theta)}$ \\
 $L(\theta)\sim$ & $r f(\theta)$ & $r f(\theta)$\\
 $\tilde c(r)\propto$ &  $r^2$ & $r^2$\\
 \hline
\end{tabular}
\caption{Large radius behaviour of the pseudo-energy, $L$--function and effective central charge in the second branch of real solutions, to first order in $r$. The function $f(\theta)$ is non-negative for rapidities $\theta\in\Theta$ in a finite subset of the real line, and negative elsewhere.}
\label{tab_asympt}
\end{table}

\paragraph{Higher order corrections}
We are interested in finding the sub-leading corrections to the asymptotic behavior $\epsilon(\theta) \sim -r f(\theta)$. We can write in full generality\footnote{In the next few formulae we temporarily display explicitly the dependence of all functions on $r$.}
\beq
    \epsilon(\theta\vert r) = -r f(\theta) + g(\theta\vert r)\;,\qquad \lim_{r\rightarrow\infty}\frac{g(\theta\vert r)}{r} = 0\;.
\eeq
As mentioned, the function $f(\theta)$ is such that
\beq
    \left\lbrace \begin{array}{l c l} f(\theta) \geq 0 & & \theta\in\Theta\;, \\ f(\theta) = 0 & \Leftrightarrow & \theta\in\partial\Theta\;, \\ f(\theta) < 0 & & \theta\in\Theta^{\perp}\;, \end{array}\qquad \Theta\subset\mathbb{R}\;. \right.
\eeq
A careful analysis of the TBA equation in the large $r$ limit, with the above conditions on the solution, reveals that $f(\theta)$ obeys the equation
\beq
    f(\theta) = - \cosh\theta + \intop_{\Theta} \varphi(\theta - \theta') f(\theta') \frac{d\theta'}{2\pi}\,,
\eeq
and that the function $g(\theta|r)$ is expanded in negative odd powers of $r$,
\beq
    g(\theta\vert r) = \sum_{\ell = 0}^{\infty}\frac{g_{2\ell+1}(\theta)}{r^{2\ell+1}}\,.
\eeq
Using this expansion one can then systematically extract from \eqref{tba_equation} the equations for the coefficient functions $g_{2\ell+1}(\theta)$. The first of these reads
\beq
    g_1(\theta) = \frac{\pi}{12}\left(\frac{\varphi(\theta - \theta')}{f'(\theta')}\right)\Bigg\vert_{\theta' \in \partial\Theta} + \intop_{\Theta} \varphi(\theta - \theta')g_1(\theta') \frac{d\theta'}{2\pi}\,.
\eeq
An interesting fact that follows from the above equations is that\footnote{One can see this by introducing the resolvent $K(\theta-\theta')$ of the kernel $\frac{1}{2\pi}\varphi(\theta-\theta')$. Then $f$ and $g_1$ are given by $f(\theta) = -\cosh\theta - \intop_{\Theta}d\theta' K(\theta-\theta')\cosh\theta'$ and $g_1(\theta) = \frac{\pi^2}{6}\left(\frac{K(\theta - \theta')}{f'(\theta')}\right)\Big\vert_{\theta'\in\partial\Theta}$. Finally, since $f(\theta)\Big\vert_{\theta\in\partial\Theta}=0$ by definition and $K(-\theta) = K(\theta)$, the identity \eqref{eq:g1_identity} follows.}
\beq
    \intop_{\Theta} \cosh\theta g_1(\theta) d\theta = -\frac{\pi^2}{6}\left(\frac{\cosh\theta}{f'(\theta)}\right)\Bigg\vert_{\theta \in\partial\Theta}\;.
\label{eq:g1_identity}
\eeq
The left-hand side of the above contributes to the constant term in the large $r$ expansion of the effective central charge \eqref{central_charge}. A careful analysis of this expansion shows that another term contributes to this order in $r$,
\beq
    \intop_{\mathbb R} d\theta \cosh\theta \log\left[1+e^{-\vert f(\theta)\vert}\right] \sim \frac{\pi^2}{6}\left(\frac{\cosh\theta}{f'(\theta)}\right)\Bigg\vert_{\theta \in\partial\Theta}\;,
\eeq
where we used the Laplace method to compute the right-hand side above. This term exactly cancels the one provided by \eqref{eq:g1_identity}, so that the effective central charge displays no constant term in its large-$r$ expansion. Sparing the reader of the uninteresting details, one can compute the contributions to order $r^{-2}$, witnessing a similar, albeit not total, cancellation of terms and arriving at the following expression
\beq
    \tilde c(r) = \frac{3}{\pi^2} r^2 \intop_{\Theta} \cosh\theta f(\theta) d\theta + \frac{1}{r^2}\left(\frac{g_1(\theta)}{f'(\theta)}-\frac{7\pi^2}{30}\frac{f''(\theta)}{f'(\theta)^3}\right)\Bigg\vert_{\theta\in\partial\Theta}+\mathcal{O}(r^{-4})\,.
\eeq
In principle it should be possible to observe the $\mathcal{O}(r^{-2})$ correction above in our numerical results, but comparison to the analytic expression is hampered by the fact that we do not have an explicit form for $f(\theta)$. We leave this point to be addressed in future work.

\subsection{Fermionic/bosonic sinh-Gordon}\label{sec:sinhGordon}

Before considering the more interesting bosonic sinh-Gordon model, we first review some aspects of the well-known fermionic theory defined by the action
\beq
\mathcal{A}=\intop d^2x\left[\frac{1}{4\pi}(\partial \phi)^2+2\mu \cosh\(2b \phi\)\right]\;.
\label{eq:shG_action}
\eeq
Here $\mu$ is a coupling constant of dimensions $[\textrm{mass}]^{2+2b^2}$, which determines the scale of the model, while $b$ is a dimensionless parameter that we take, for reasons that will be clarified momentarily, to lie in the interval $b\in[0,1]$. The action \eqref{eq:shG_action} is amenable to various interpretations. First, we can understand it as defining a free Gaussian theory deformed by the relevant operator $\cosh\(b \phi\)$ with negative dimension\footnote{The reader might object that a perturbation by an operator of negative dimension destroys unitarity. This is actually not the case for sinh-Gordon theory, due to the presence of fields with non-vanishing vacuum expectation values which alter the two-point function's UV behaviours from their expected CFT ones. A discussion of this point can be found in \cite{Zamolodchikov:1990bk}.} $\Delta = -b^2$. Another possibility is to consider \eqref{eq:shG_action} as a perturbation of Liouville CFT by the operator $e^{-2b\phi}$, again of negative dimension $\Delta = -b^2$. A further possible interpretation, presented in \cite{Babelon:1990bq}, sees the sinh-Gordon model as a conformal affine $\mathfrak{sl}_2$ Toda field theory with spontaneously broken conformal symmetry. Here we will adopt the first point of view and take sinh-Gordon theory to be a relevant RG flow from a free bosonic theory in the UV to a trivial IR fixed point. As is well-known, this model is integrable, its spectrum consisting of a single neutral particle of mass\footnote{The particle mass is related to the parameter $\mu$ in the action \eqref{eq:shG_action} by the famous formula \cite{Zamolodchikov:1995xk} $$\pi\mu\frac{\Gamma\left(b^2\right)}{\Gamma\left(1-b^2\right)} = \left[\frac{m}{8\sqrt{\pi}} a^a\left(1-a\right)^{1-a}\Gamma\left(\frac{a}{2}\right)\Gamma\left(\frac{1-a}{2}\right)\right]^{2+2b^2}\;.$$} $m$ subject to a factorised two-body scattering with amplitude \cite{Vergeles:1976ra}
\beq
    S^\text{\,f}_\text{shG}(\theta)=\frac{\sinh\theta-i\sin\(\pi a\)}{\sinh\theta+i\sin\(\pi a\)}\,,
\la{S_fshG}
\eeq
where $a$ and $b$ are related by
\beq
    a = \frac{b^2}{1+b^2}\;.
\eeq
Notice that the $S$-matrix \eqref{S_fshG} is invariant under the weak-strong duality $b\rightarrow 1/b$, or equivalently $a\rightarrow 1-a$. This is the reason why we restricted the parameter $b$ to the interval $b\in[0,1]$.

Given the $S$-matrix \eqref{S_fshG} we can compute the effective central charge via the TBA from \eqref{central_charge}. In this case, the TBA equation can be solved by a simple iterative routine as described in section~\ref{sec:numerical_methods}, and figure~\ref{fig_shG_ceff} shows the resulting curves $\tilde c(r)$ for different values of the coupling $a$.
\begin{figure}[h!]
\centering
\includegraphics[width=.8\textwidth]{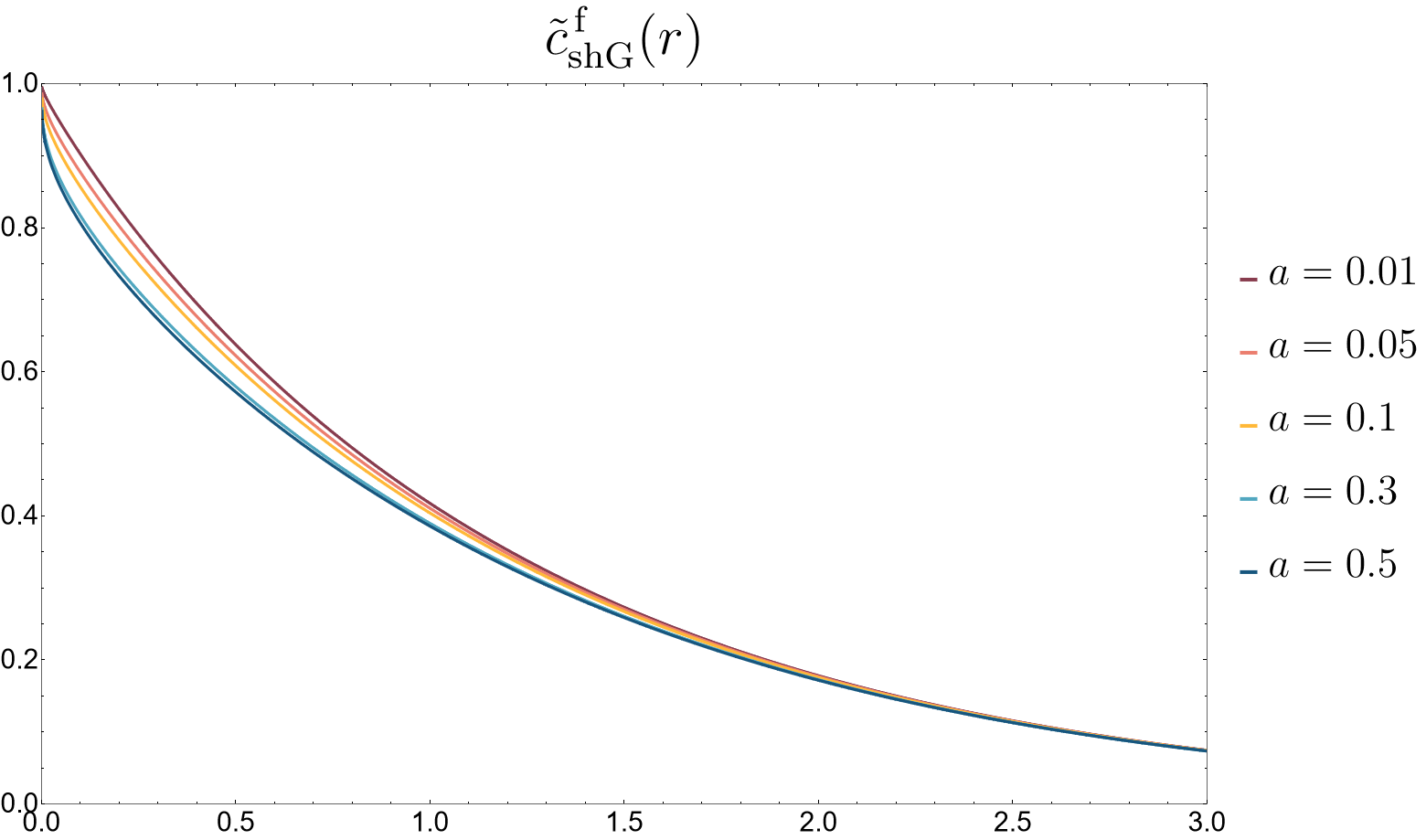}
\vspace{0.3cm}
\caption{Effective central charge of the fermionic sinh-Gordon model at various values of the coupling $a$. Numerical solutions were computed using the iterative algorithm described in section~\ref{sec:numerical_methods} with $10^5$ points in the interval $\theta \in [-10,10)$.} 
\la{fig_shG_ceff}
\end{figure}
In particular, we see that there is a single branch of solutions displaying the expected interpolation between the IR and UV behaviours,
\beq
    \lim_{r\rightarrow \infty} \tilde c(r) = 0\;,\qquad 
    \lim_{r\rightarrow 0} \tilde c(r) = 1\;.
\eeq
The above result for the ultraviolet limit of the effective central charge can be obtained following the lines illustrated in section~\ref{sec:elliptic_model}. The peculiarity with this case is that the solution is such that $L\rightarrow\infty$ and $\epsilon\rightarrow-\infty$. The dilogarithm trick can nonetheless still be applied and at the end of the day one sees that
\beq
    \lim_{r\rightarrow 0} \tilde c(r)=  \lim_{\epsilon\rightarrow -\infty}\, \frac{6}{\pi^2}\,\text{L}\left(\frac{1}{1+e^{\epsilon}}\right) = 1\;.
\eeq

Sub-leading effects were computed in \cite{Zamolodchikov:1992ulx} and \cite{Zamolodchikov:2000kt}, and include soft logarithmic corrections of the form
\beq
     \lim_{r\rightarrow 0} \tilde c(r)= 
     1-\frac{3\pi^2}{2}\frac{a(1-a)}{\left(\log r\right)^2} + \mathcal{O}\left(\log r\right)^{-3}\;.
\eeq

~

We now turn to the bosonic sinh-Gordon model defined through the $S$-matrix
\beq
S^\text{\,b}_\text{shG}(\theta)=-\frac{\sinh\theta-i\sin\(\pi a\)}{\sinh\theta+i\sin\(\pi a\)}\,,
\la{S_bshG}
\eeq
which differs from the usual sinh-Gordon one by an overall sign. We call this the {\it bosonic} sinh-Gordon model since \eqref{S_bshG} is of bosonic type, \textit{i.e.} $S^\text{b}_\text{shG}(0)=1$. Even though there is no associated Lagrangian or known physical realization for this $S$-matrix, we take the point of view in which we explore the space of quantum field theories by analyzing consistent $S$-matrices. In this sense, the $S$-matrix \eqref{S_bshG} along with the assumption of elastic scattering provides the definition of the theory. 

\begin{figure}[h!]
\centering
\includegraphics[width=.8\textwidth]{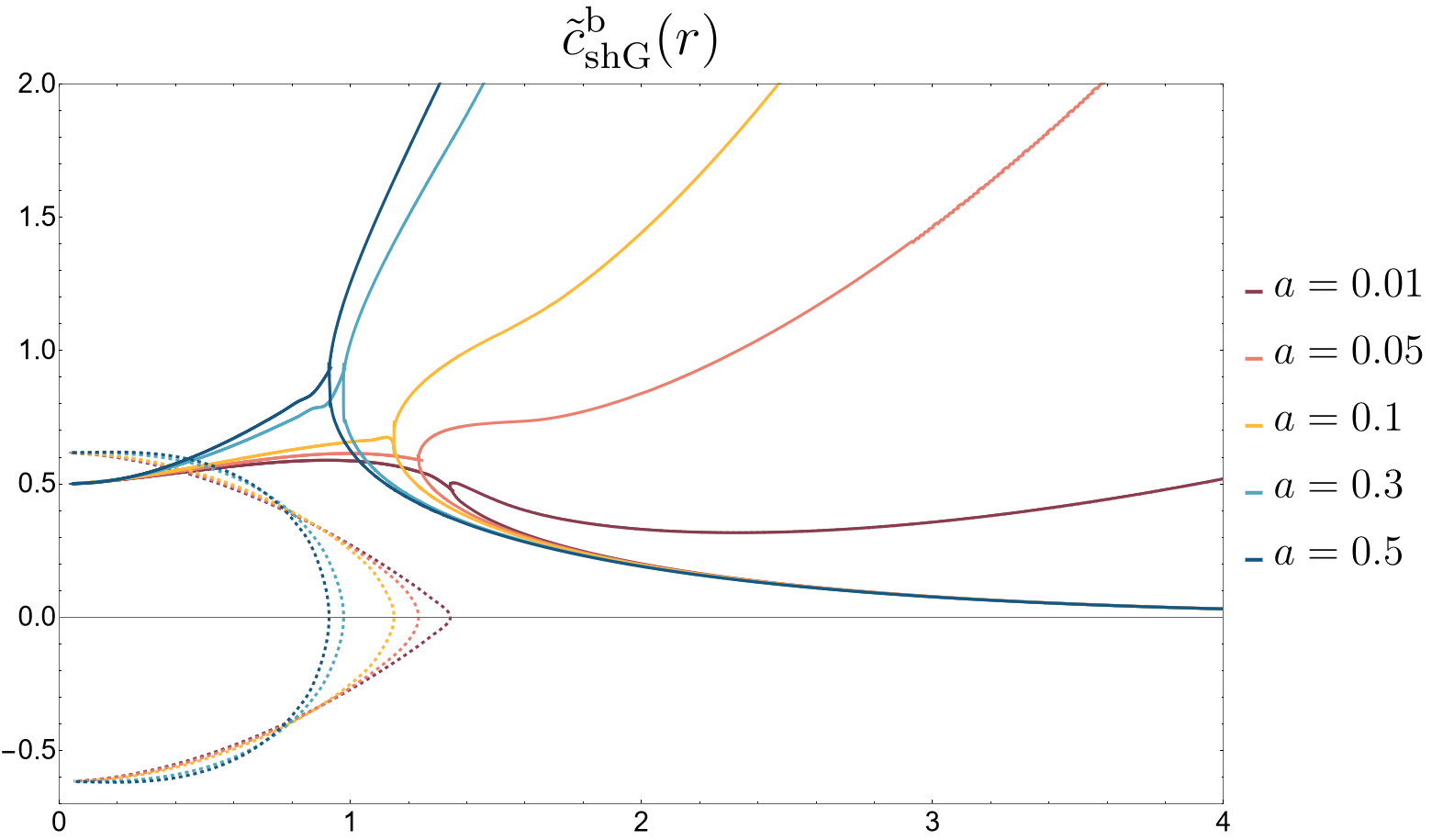}
\vspace{0.3cm}
\caption{Effective central charge for the bosonic sinh-Gordon model at various values of the coupling $a$. Above a critical radius $r_c$ there are two branches of real solutions, whereas below this radius there is a pair of complex-conjugate solutions with non-vanishing imaginary parts (dashed lines). The complex effective central charge in the ultraviolet $\tilde c(r=0)\approx 0.5\pm0.617 \ii$ can be computed from the plateau solutions as explained in the main text.  Numerical solutions were computed using the pseudo-arclength continuation method described in section~\ref{sec:numerical_methods} with $500$ points in the interval $\theta \in [-6,6)$.}
\la{fig_bshG_ceff}
\end{figure}

The seemingly harmless change of overall sign in the $S$-matrix \eqref{S_bshG} has profound consequences for the TBA. As figure~\ref{fig_bshG_ceff} shows, there is now a critical radius $r_c$ where we encounter a bifurcation. Above this critical radius we find two branches of real solutions, whereas below it there is a pair of complex conjugate solutions. From the point of view of the real solutions, the critical radius is a turning point in which the two possible branches, \textit{i.e.} the one with free asymptotic behaviour and the second one with asymptotics described in section~\ref{sec:asymptotic_analysis}, merge. Let us focus on these solutions first. 

The existence at large radius of two, and only two, real branches is compatible with the analysis of \cite{Camilo:2021gro} for generalized $\TTB$ deformations. That is, for any coupling $a\in(0,1/2]$ the $L_1$ norm of the kernel is equal to 1, and for the bosonic TBA equation this is enough to allow for a second real branch. The branch with lower effective central charge can be found numerically with the usual iterative method starting from the free solution in the IR. To get very close and eventually pass the critical radius, we need however to use the pseudo-arclength continuation method described in section~\ref{sec:numerical_methods}. Close to the turning point the curve is well approximated by a square root, $\tilde c\approx \alpha \sqrt{r-r_c}$ for some constant $\alpha$, reminiscent of the situation  in $\TTB$ deformed theories. The second branch behaves like $\tilde c\sim r^2$ at large radius, as expected.

\begin{figure}[httt!]
\centering
\begin{subfigure}[t]{0.75\textwidth}
\centering
\includegraphics[width=\textwidth]{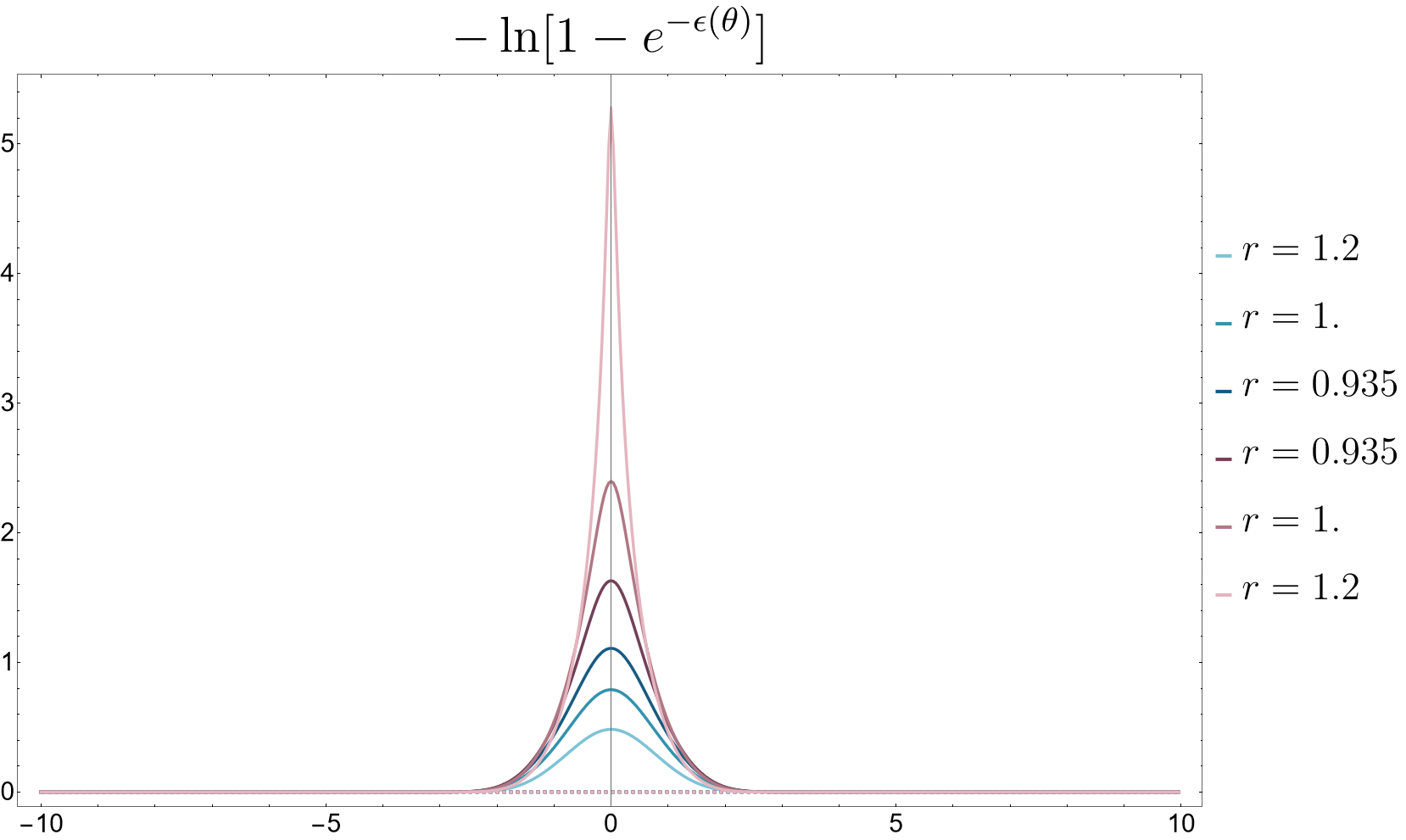}\vspace{0.3cm}
\end{subfigure}
\hfill
\begin{subfigure}[t]{0.85\textwidth}
\centering
\includegraphics[width=\textwidth]{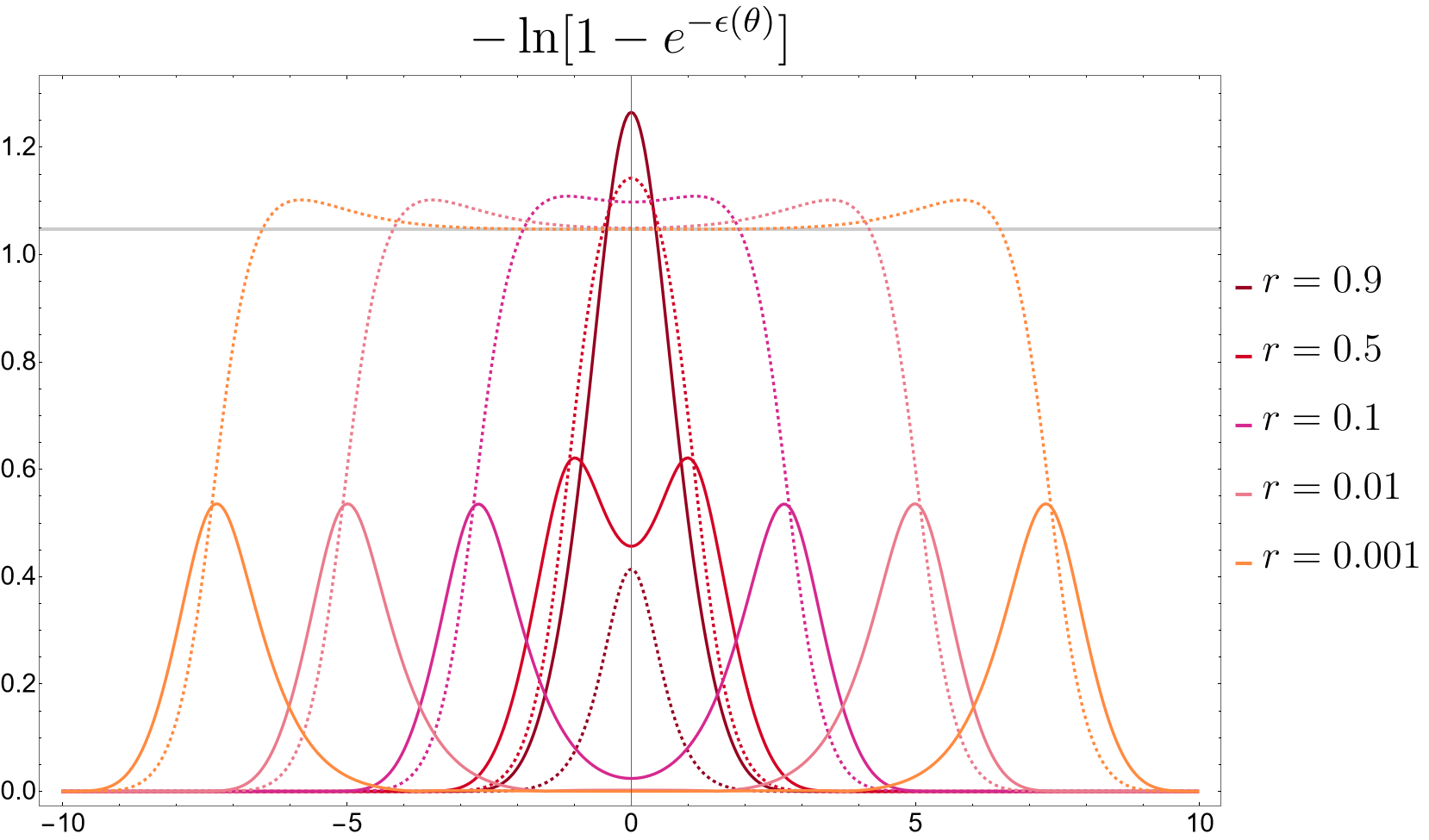}\vspace{0.3cm}
\end{subfigure}
\caption{$L(\theta)=-\ln\[1-e^{-\epsilon(\theta)}\]$ functions for the bosonic sinh-Gordon model with $a=1/2$ at different points along the curve of figure~\ref{fig_bshG_ceff}. The top figure shows the solutions for $r>r_c$ where we have two real branches ($r_c\approx0.93$ for $a=1/2$). In blue we have the usual branch with free asymptotic behaviour $\epsilon(\theta)\sim r \cosh(\theta)$; and in purple the second branch with asymptotic behaviour as in table~\ref{tab_asympt}. In the bottom figure we have the complex solutions below $r_c$, which form a purely imaginary plateau (dashed lines) as $r\rightarrow0$.}
\la{fig_bshG_Ls}
\end{figure}

Note that the free theory limit $a\rightarrow0$ is not smooth, since the $L_1$ norm of the kernel changes discontinuously from 1 for any $a\in(0,1)$ to 0 for $a=0,1$. We find numerically that as the coupling decreases the critical radius grows, see right panel of figure~\ref{fig_rc}, while the second real branch approaches the first one at small scales, see figure~\ref{fig_bshG_ceff}. Of course, both branches later separate as is required by their corresponding $\tilde c\sim 0$ and $\tilde c \sim r^2$ asymptotics. The $L$--functions are plotted for different radii in figure~\ref{fig_bshG_Ls} (top) and are again consistent with the expected asymptotic behaviour summarized in table~\ref{tab_asympt}.

Below $r_c$ solutions to the TBA equation become complex. As mentioned in section~\ref{sec:numerical_methods}, to find these solutions it suffices to apply the pseudo-arclength continuation method allowing pseudo-energies to have real and imaginary parts, introducing small perturbations close to the critical radius to switch from the real to the complex branch. Note that once an imaginary part has been acquired by the pseudo-energy the perturbations can be turned off as the corresponding family of solutions is continued further. 

As seen in figure~\ref{fig_bshG_ceff}, the effective central charge reaches in the ultraviolet limit $r\to0$ a pair of complex-conjugate values that is independent of the coupling $a$.
We can verify numerically that the $L$--functions form a plateau at small radii, as shown in figure~\ref{fig_bshG_Ls}~(bottom), so that the dilogarithm trick outlined in section~\ref{sec:elliptic_model} is applicable. Since the integral of the kernel is $\Omega = 1$ we find two complex solutions,
\beq
\epsilon=+ \ln\(1- e^{-\epsilon}\) \quad\implies\quad \epsilon=\pm\, \ii\pi/3.
\eeq
resulting in $\tilde c^\text{ b}_\text{shG}(0) \approx 0.5\pm0.617 i$ which matches the pair of complex conjugate values found numerically, see figure~\ref{fig_bshG_ceff}. 
The fact that these values are the same for any coupling can thus be traced back to the $L_1$ norm of the kernel being independent of the coupling $a$.
%

\subsection{Fermionic/bosonic elliptic sinh-Gordon}

We now turn to the elliptic sinh-Gordon model. As reviewed in section~\ref{sec:elliptic_model} the model has an $S$-matrix with periodicity $S(\theta+\Tl)=S(\theta)$, controlled by the modulus $l$. There is an infinite set of resonances for any $l\in(0,1)$  and no bound states. We can therefore immediately infer that two real branches are allowed in the TBA, and thus expect qualitatively similar behavior to the bosonic sinh-Gordon model, including the existence of a critical radius representing a turning point. However, this needs to be checked numerically since the arguments we have access to at the moment only constrain the asymptotic behavior of the solutions and tells us nothing about their local properties.

\begin{figure}[h!]
\centering
\includegraphics[width=.8\textwidth]{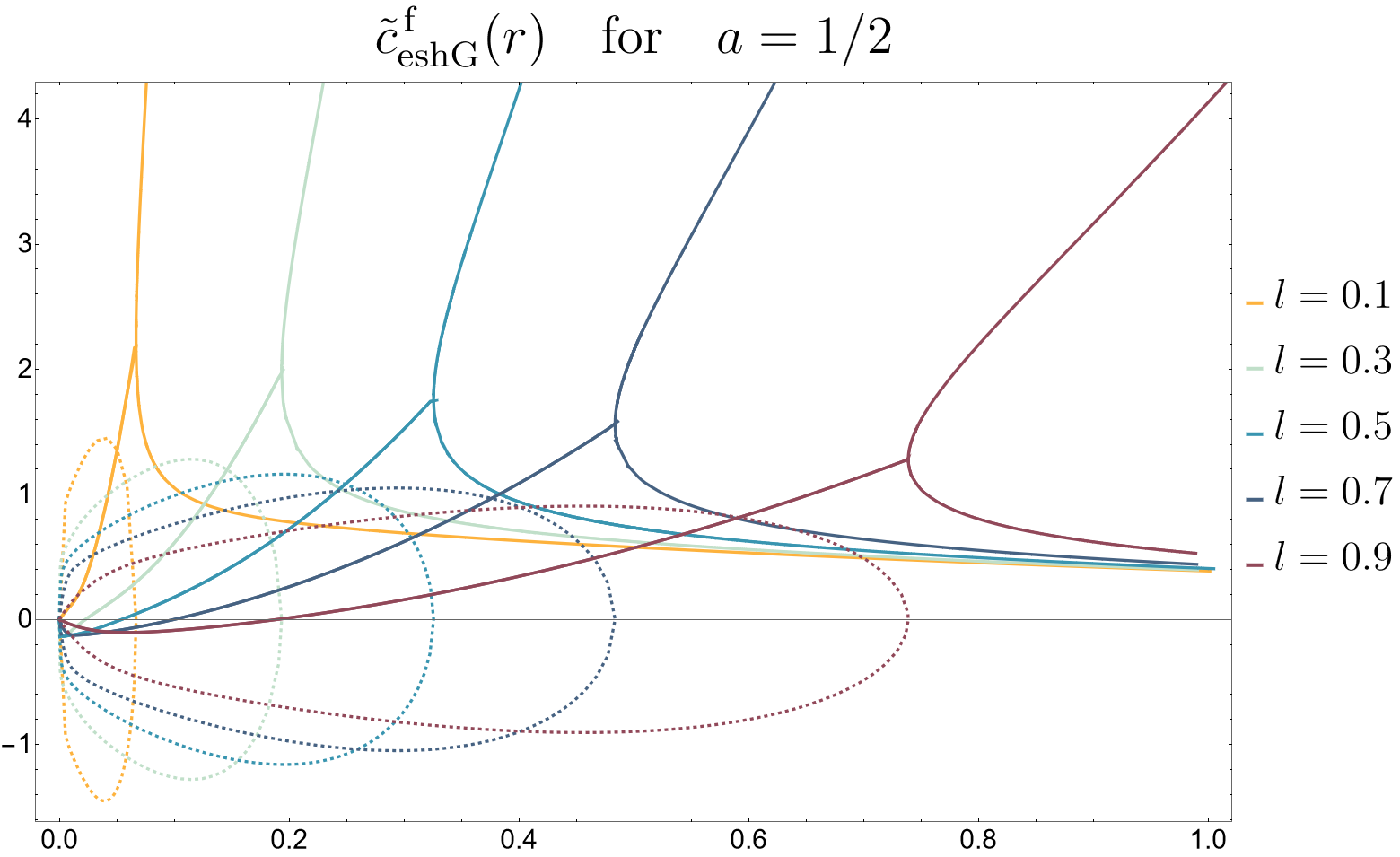}\\[5pt] \includegraphics[width=.8\textwidth]{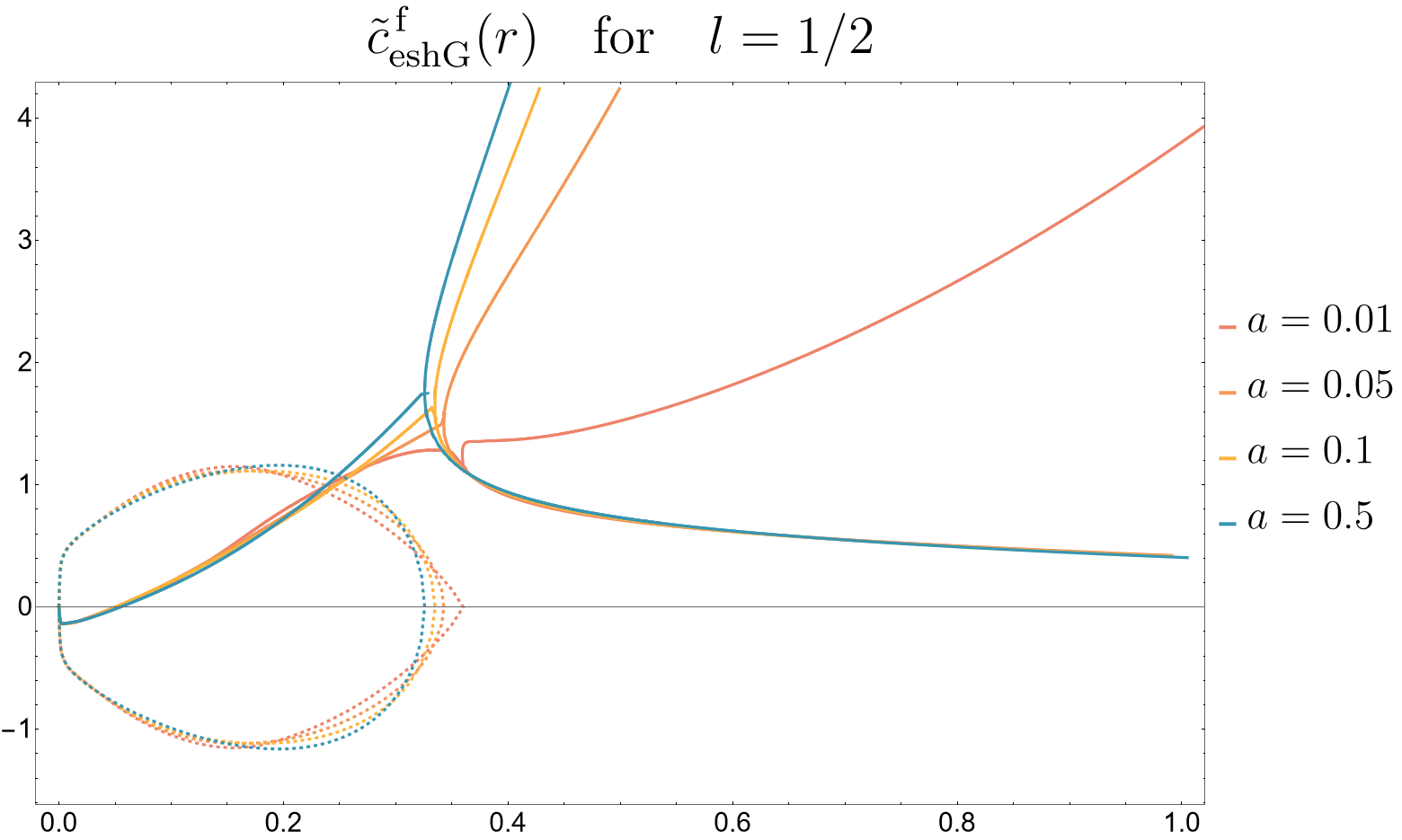}\vspace{0.3cm}
\caption{Effective central charge for the fermionic elliptic sinh-Gordon model. Plots for various values of $l$ with fixed $a=1/2$ (top), and various values of $a$ with fixed $l = 1/2$ (bottom). Above a critical radius $r_c$ there are two branches of real solutions, whereas below this radius there is a pair of complex-conjugate solutions with non-vanishing imaginary parts (dashed lines). The effective central charge in the ultraviolet $\tilde c(r=0) = 0$ can be computed from the plateau solutions as explained in the main text.  Numerical solutions were computed using the pseudo-arclength continuation method described in section~\ref{sec:numerical_methods} with $N=200$ and $M=2$.}
\label{fig_ferm_ellshG}
\end{figure}

\begin{figure}[h!]
\centering
\includegraphics[width=.8\textwidth]{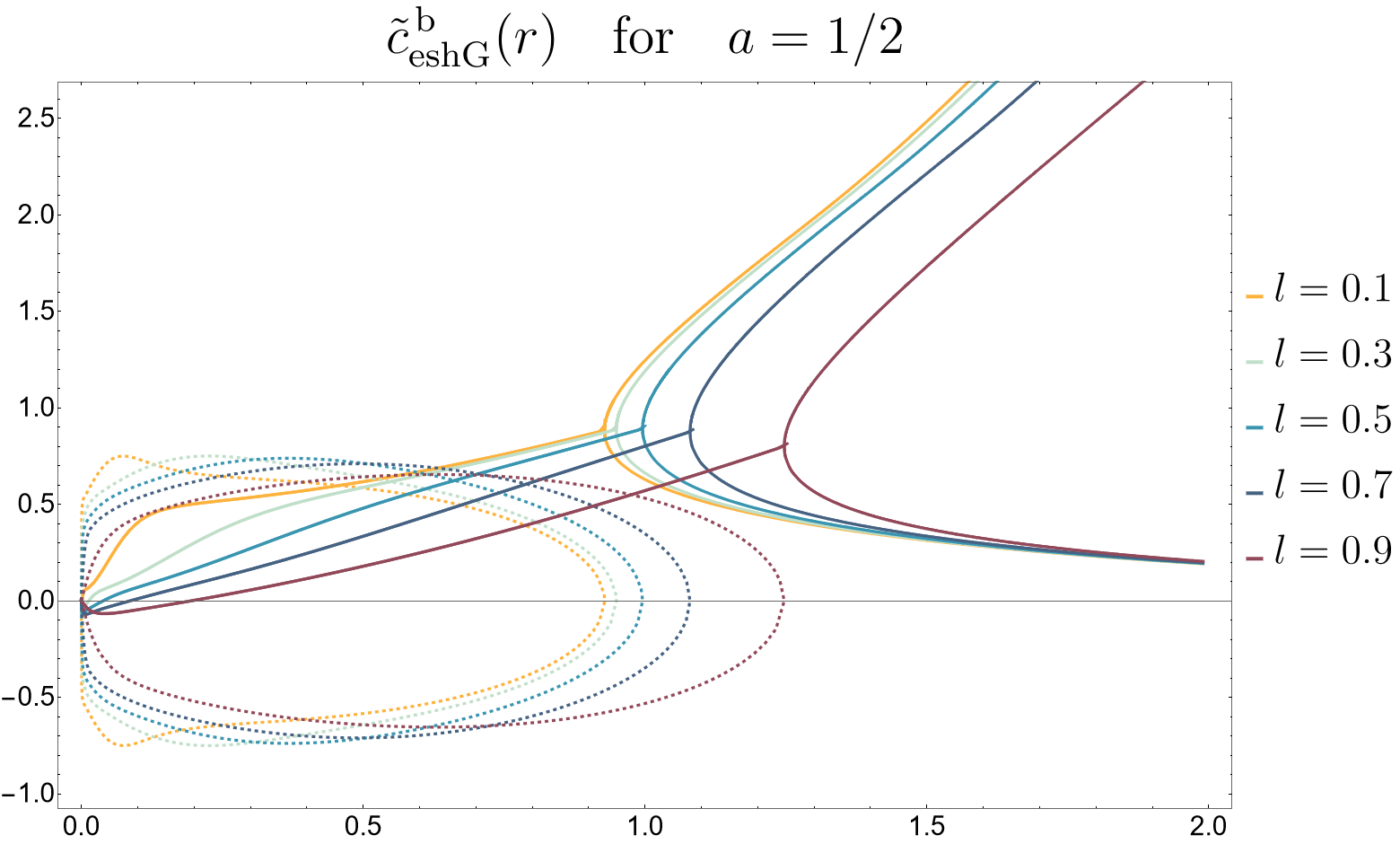}\\[5pt] \includegraphics[width=.8\textwidth]{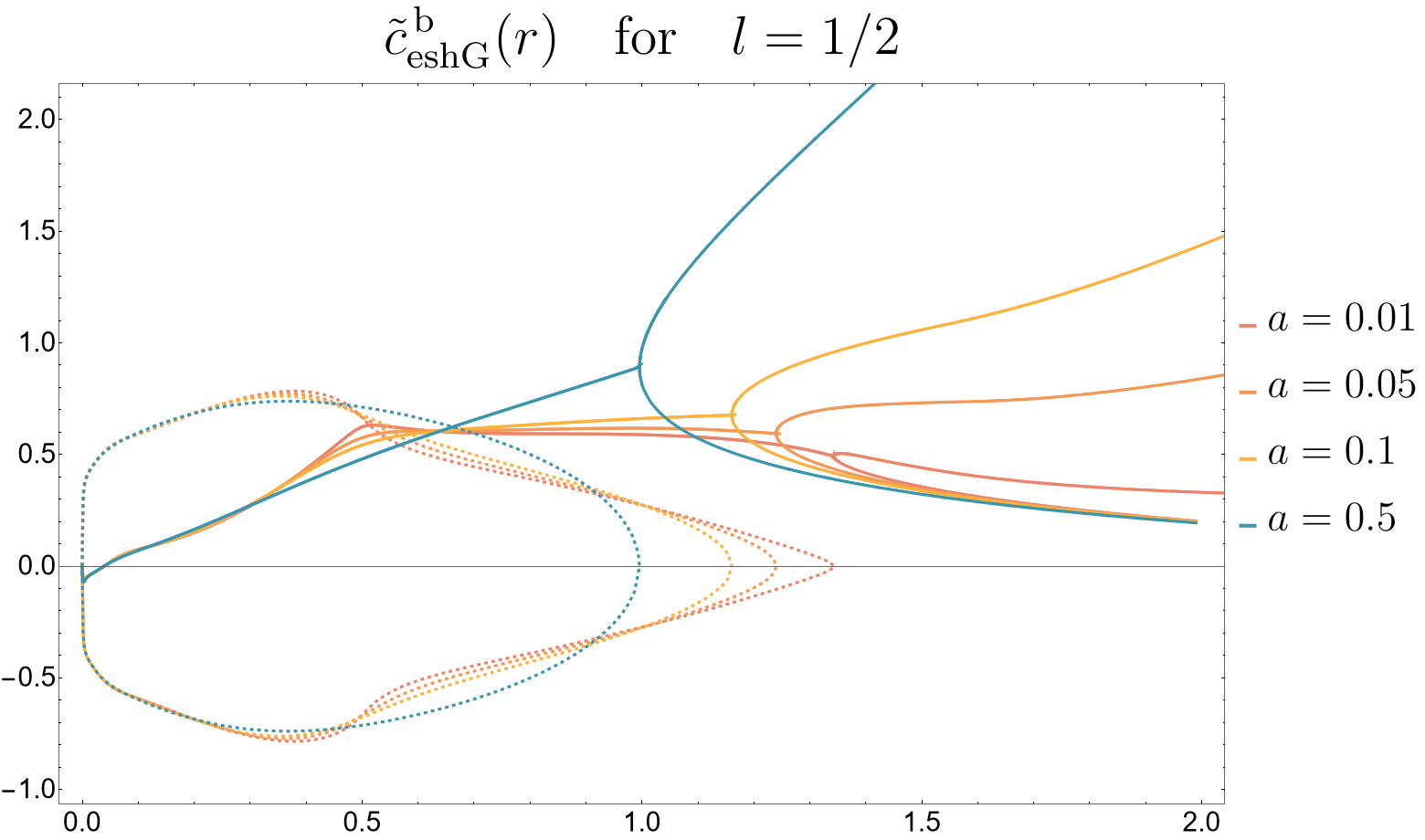}\vspace{0.3cm}
\caption{Effective central charge for the bosonic elliptic sinh-Gordon model. Plots for various values of $l$ with fixed $a=1/2$ (top), and various values of $a$ with fixed $l = 1/2$ (bottom). Above a critical radius $r_c$ there are two branches of real solutions, whereas below this radius there is a pair of complex-conjugate solutions with non-vanishing imaginary parts (dashed lines). The effective central charge in the ultraviolet $\tilde c(r=0) = 0$ can be computed from the plateau solutions as explained in the main text.  Numerical solutions were computed using the pseudo-arclength continuation method described in section~\ref{sec:numerical_methods} with $N=200$ and $M=2$.}
\la{fig_bosonic_ellshG}
\end{figure}

\begin{figure}[h]
\centering
\begin{tabular}{cc}
\includegraphics[width=.475\textwidth]{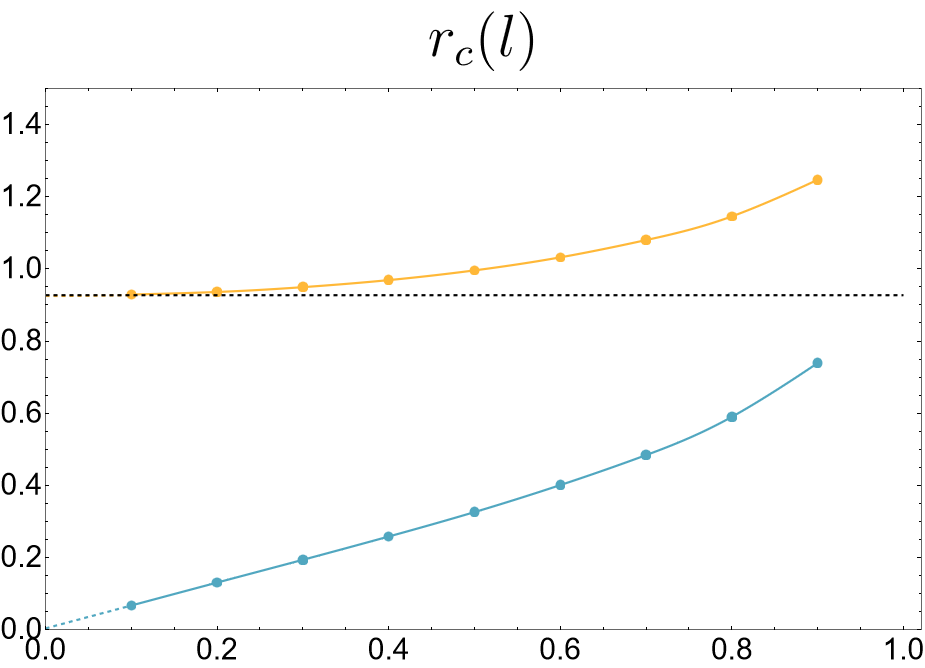}& 
\includegraphics[width=.475\textwidth]{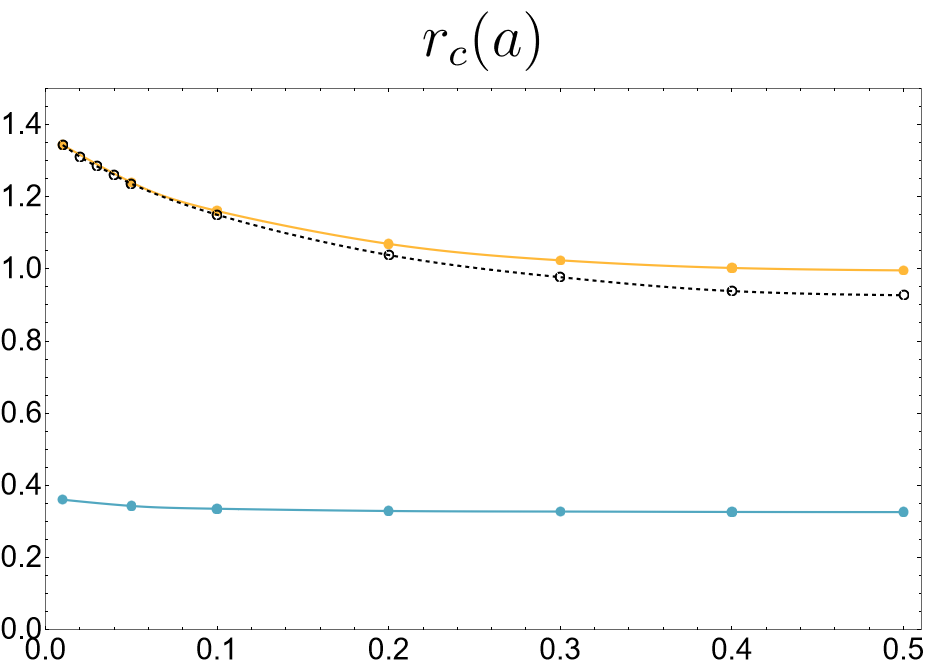}
\end{tabular}
\caption{Critical radius as a function of the modulus $l$ (left, $a = 1/2$) and of the coupling $a$ (right, $l = 1/2$). Results for the standard bosonic, elliptic fermionic and elliptic bosonic sinh-Gordon models are shown in dashed black and solid blue and orange, respectively. On the left panel the colored dashed lines show extrapolations for $l\to0$: the fermionic case linearly approaches $r_c(l=0) = 0$, whereas the bosonic case has $r_c(l) = r_c(l=0) + \frac{l^2}{4(1-l)^{1/5}}$ with $r_c(l=0) \approx 0.995$, which fits all the datapoints and matches at $l=0$ the critical radius obtained for the bosonic sinh-Gordon model. On the right panel, the critical radii of the bosonic standard and elliptic sinh-Gordon models are very similar for small $a$, displaying more significant differences as $a\approx1/2$, while staying always above the critical radius of the fermionic elliptic model.}
\label{fig_rc}
\end{figure}

As it turns out, the elliptic sinh-Gordon model has two real branches in both its fermionic and bosonic flavors, see figures~\ref{fig_ferm_ellshG} and \ref{fig_bosonic_ellshG}.\footnote{In a previous attempt to solve the TBA equation for this model \cite{Castro-Alvaredo:2001yqz}, the authors mention poor convergence of their iterative method at small radii.} We find that the critical radius where the turning point is found is always greater in the bosonic than in the fermionic case, see figure~\ref{fig_rc}. Moreover, the critical radius decreases as we decrease the modulus $l$ (left panel). In the fermionic case this was expected because in the limit $l\rightarrow0$ we should recover the results for the fermionic sinh-Gordon model, which does not have two branches. In the bosonic case, the critical radius approaches in the $l\to0$ limit the corresponding value in the bosonic sinh-Gordon model, \textit{cf.} rightmost point of the dashed black curve on the right panel.

The dependence of the critical radius on the coupling $a$ is milder than that on the modulus $l$, as can be seen in the right panel of figure~\ref{fig_rc}. We find $r_c$ decreases with $a$,  
 but this is only noticeable when $a \approx 0$. Similarly to the bosonic sinh-Gordon case, the $a\to0$ limit is not smooth, which again traces back to the fact that the norm of the kernel changes discontinuously at $a = 0$.
 
There is a surprising property of the family of real solutions, namely the appearance of an algebraic relation between the kernel $\varphi_{a,l}(\theta)$ and the convolution term in the TBA equation.  We leave this discussion to appendix~\ref{app:algebraic_relation}, and now turn instead our attention to the complex solutions.

Evaluating the augmented Jacobian determinant \eqref{augmented_jacobian} along the family of real solutions, we are able to see that the turning point is actually a simple bifurcation point for the complexified theory. Indeed, following the real branch from the standard large $r$ regime with $\epsilon(\theta) \approx r \cosh(\theta)$ to the non-standard regime with $\tilde c(r) \sim r^2$, we have $J = -1$ in the lower branch, \textit{i.e.} before the critical point, and $J = 1$ afterwards. The precise location of the critical point can thus be pinpointed using binary search on the pseudo-arclength step size $\Delta s$. Below the critical radius $r_c$ we have a pair of complex conjugate solutions, which in this case have $\tilde c \to 0$ as $r\to0$. Interestingly, the UV behavior is only observed very close to $r = 0$.

To better understand the $\tilde c\rightarrow0$ behaviour, we consider $k$--CDD factors approximating the elliptic sinh-Gordon $S$-matrix, and take the $k\rightarrow\infty$ limit. That is, we take the $S$-matrix
\beq
S^\text{\,f,b}_{k\text{CDD}}(\theta)=\pm\prod\limits_{j=-\lfloor k/2\rfloor}^{\lfloor k/2\rfloor}\frac{\sinh(\theta)-\sinh(\ii\pi a+ j\, T_l)}{\sinh(\theta)+\sinh(\ii\pi a+ j\,T_l)}\,,
\eeq
and analyze it as before with the help of the TBA. In practice, the differences between the elliptic and the $k$--CDD models above is that the corresponding kernels start to differ for $|\theta|\sim\lfloor k/2\rfloor T_l$. Since we are truncating all numerical integrations, we need to make sure to include the tails of the $k$--CDD kernel at large $\theta$. Consequently, deviations are only noticeable for small $r$.

Guided by experimental numerical evidence for several $k$'s, let us assume the associated TBA has a turning point and the complex solutions reach a plateau as $r\rightarrow0$. Since the $L_1$ norm of the kernel counts the number of resonances, these constant solutions satisfy equation \eqref{r_to_0_tba} with $\Omega = k$.
There is a simple geometric depiction in terms of the $Y$--functions $Y(\theta)=e^{-\epsilon(\theta)}$, for which the equations read
\beq\la{Yconst}
Y=(1\pm Y)^{\pm k}\,, 
\eeq
where the upper (lower) signs refer as usual to the fermionic (bosonic) TBA.

\begin{figure}[h!]
\centering
\begin{subfigure}[t]{0.495\textwidth}
\centering
\includegraphics[width=\textwidth]{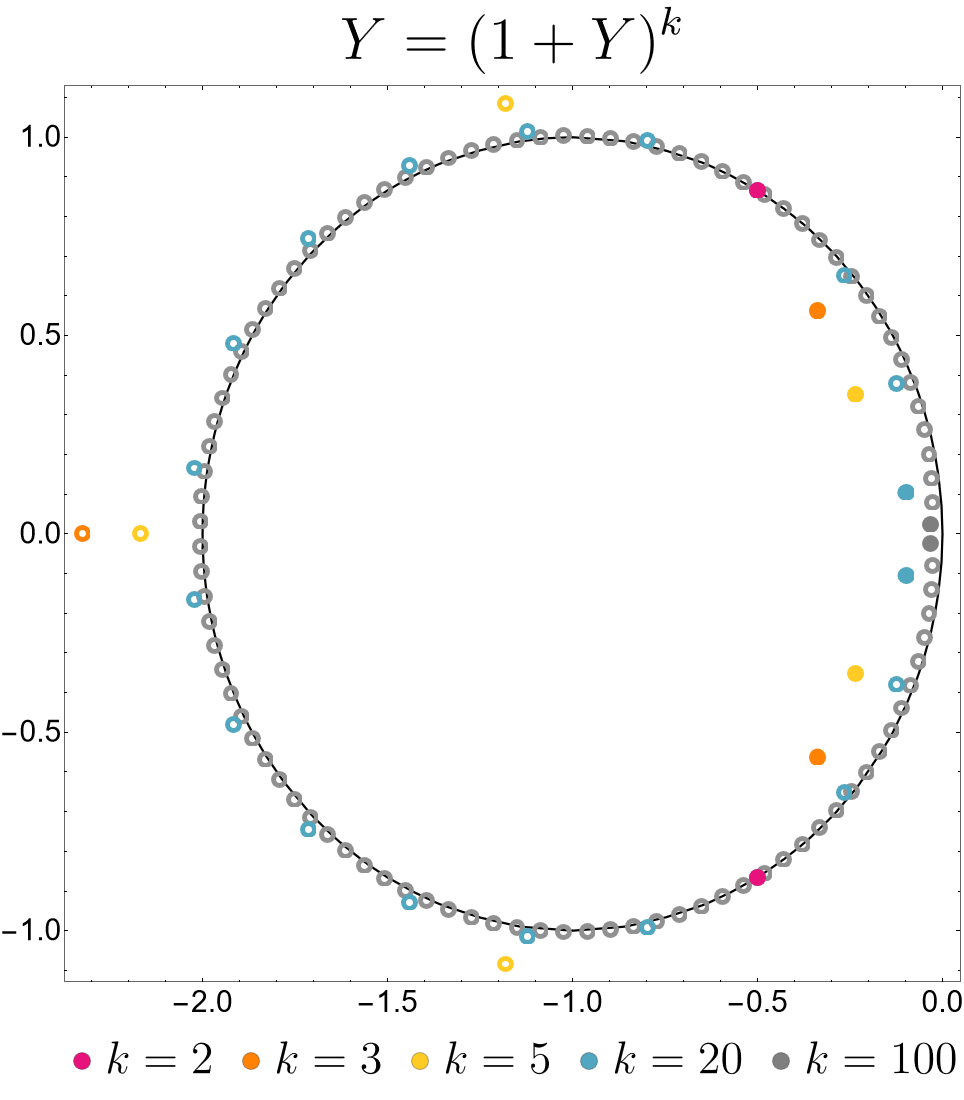}\vspace{0.3cm}
\end{subfigure}
\hfill
\begin{subfigure}[t]{0.495\textwidth}
\centering
\includegraphics[width=\textwidth]{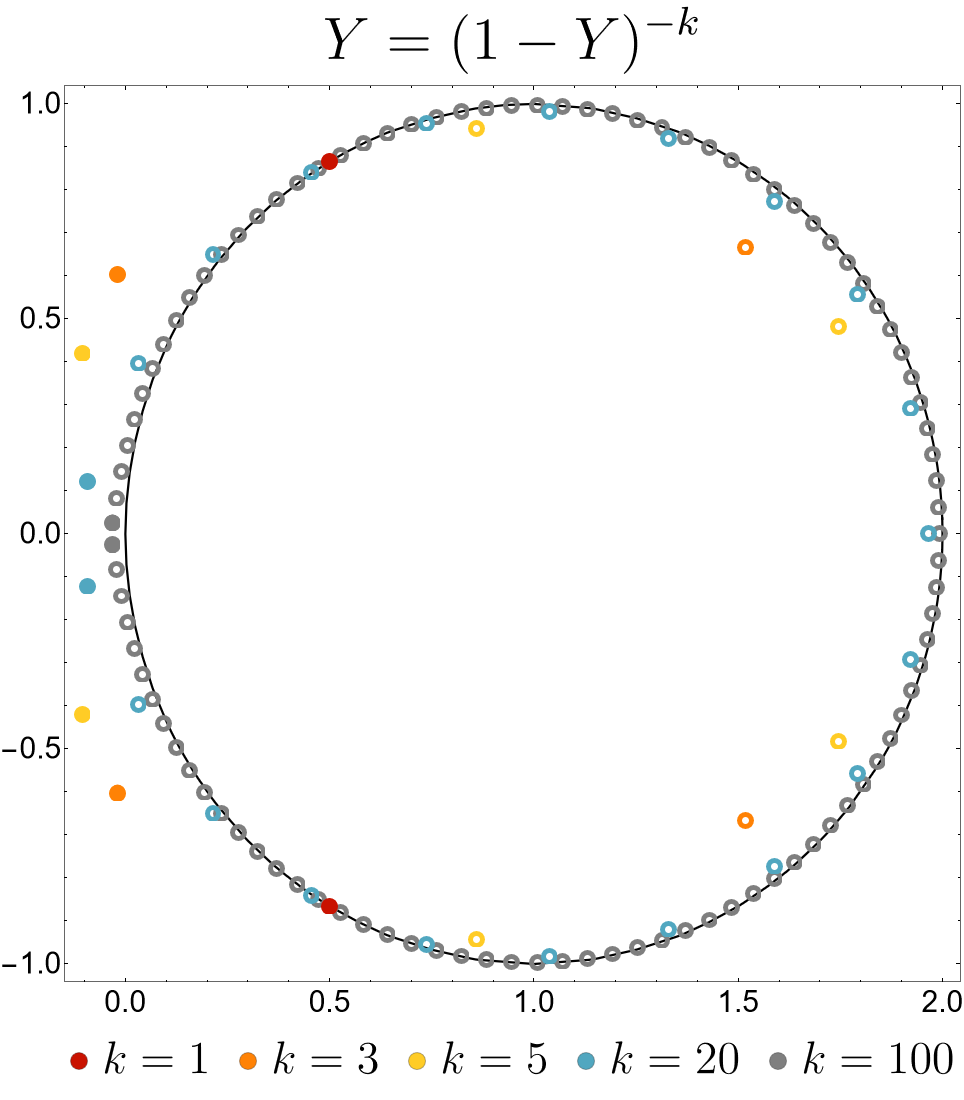}\vspace{0.3cm}
\end{subfigure}
\caption{Constant solutions to equations \eqref{Yconst} in the complex plane, where $k$ labels the number of resonances and $Y$ are the constant $Y$--functions $Y=e^{-\epsilon}$ in the ultraviolet. There are $k$ solutions to the fermionic equations (left) and $k+1$ to the bosonic ones (right). The filled circles identify the pair of solutions chosen by the TBA. In the $k\rightarrow\infty$ limit, the solutions condense into the circles centered at $Y=\mp1$, depicted in black. In this limit the TBA gives $Y\rightarrow0$ for both bosonic and fermionic equations, resulting in $\tilde c=0$ in the ultraviolet.}  
\label{fig:Y_solns}
\end{figure}
In figure~\ref{fig:Y_solns} we plot solutions to the above equation for different values of $k$. In general, there are $k$ solutions to the fermionic equation and $k+1$ to the bosonic one.\footnote{This counting is in agreement with the map between bosonic and fermionic TBA in \eqref{eq:map_bos}.} As $k$ increases, the solutions condense into circles of unit radii centered around $Y=\mp1$, see figure~\ref{fig:Y_solns}. The TBA, however, picks only two of these $\cO(k)$ solutions. Which ones? The compelling answer is: the solutions which minimize $|\tilde c|$ as computed with Rogers' dilogarithm \eqref{c_dilog}. In terms of the plots in figure~\ref{fig:Y_solns}, the TBA chooses the pair of solutions closer to $Y=0$. 
The analysis above is confirmed by our numerical solution to the TBA equations for the first few non-trivial values of $k$, namely $k\leq7$ both in bosonic and fermionic equations. 
Note that the norm of the kernel is independent from the parameters entering the CDD factors. Moreover, since we can change these parameters continuously and a different solution to \eqref{Yconst} would involve a discontinuous change, we expect the TBA to choose the solutions closer to $Y=0$ no matter the resonance locations.\footnote{For instance, for the 2-CDD model it was shown in \cite{Camilo:2021gro} that the TBA has a turning point for any values of the parameters entering the $S$-matrix, so that the associated complex solutions would agree in the ultraviolet with figure~\ref{fig:Y_solns}.} This intuition should be nonetheless confirmed with a detailed numerical analysis.

\begin{figure}[ht!]
\centering
\includegraphics[width=.6\textwidth]{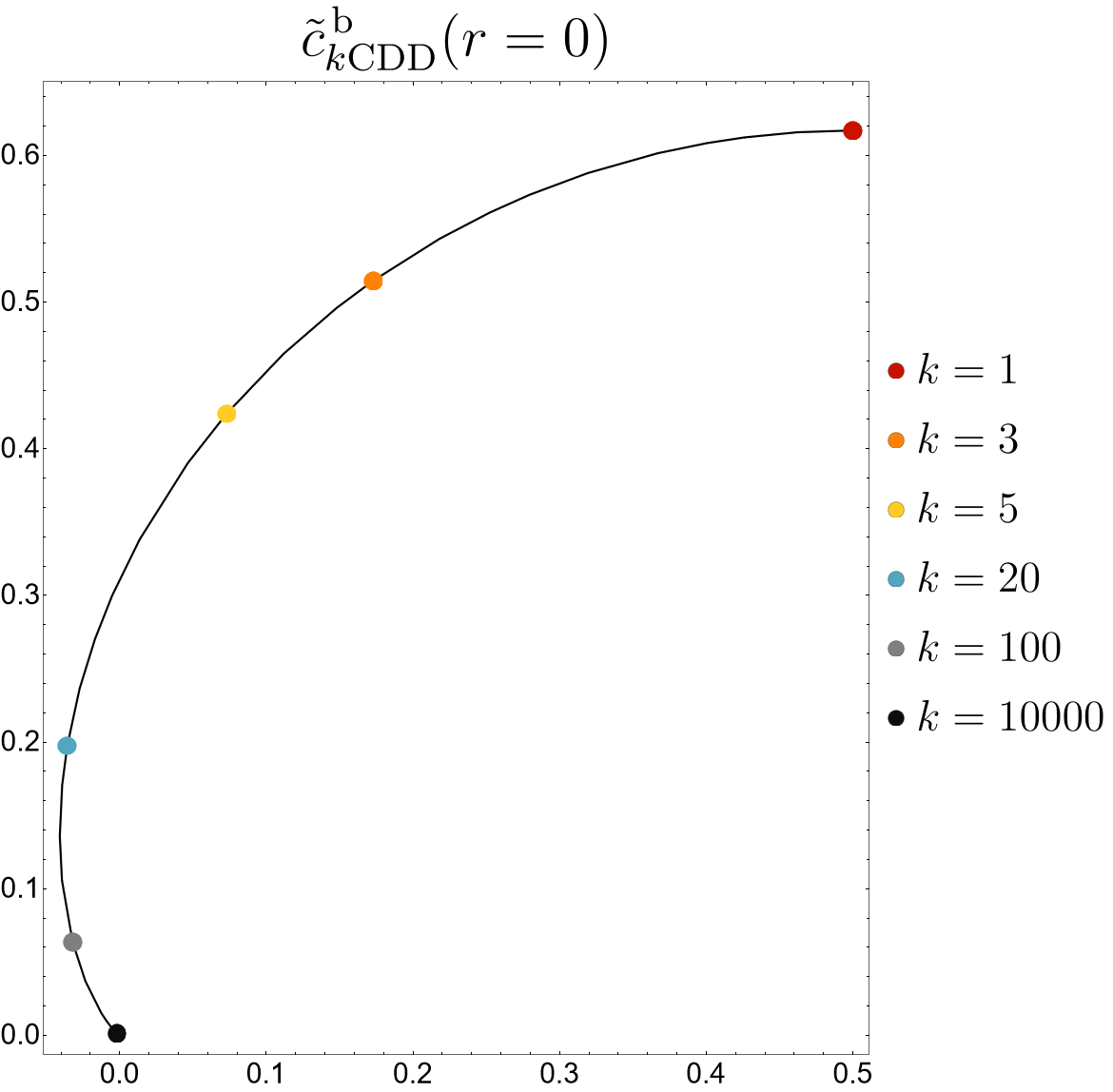}\vspace{0.3cm}
\caption{Effective central charge $\tilde c(r=0)$ in the complex plane as we increase the number $k$ of resonances in the theory, computed for the bosonic TBA with the constant solutions of figure~\ref{fig:Y_solns} and Rogers' dilogarithm formula \eqref{c_dilog}. We show only one of the complex conjugate pairs for simplicity. The corresponding curve for fermionic theories is the same provided we map $k\rightarrow k+1$. The first point $k=1$ in red matches the bosonic sinh-Gordon model value, \textit{cf.} figure~\ref{fig_bshG_ceff}. In the limit with infinite resonances $k\rightarrow\infty$ the complex effective central charge goes to zero as we find in the elliptic sinh-Gordon models, see figures~\ref{fig_ferm_ellshG} and \ref{fig_bosonic_ellshG}.}
\label{fig:c_kCDD}
\end{figure}

In figure~\ref{fig:c_kCDD} we show the ultraviolet effective central charge in the complex plane as we increase the number of resonances in the theory. The curve starts at the bosonic 1-CDD model as the bosonic sinh-Gordon model in section~\ref{sec:sinhGordon} and goes to zero as the number of resonances goes to infinity as in the elliptic sinh-Gordon models. 


\section{Bound state explorations: deformed Minimal Models}\label{sec:minimal_models}

In this section we initiate the study of the CDD deformations of models with more than one stable particle. We choose as the undeformed theory the well-known $\Phi_{1,3}$ integrable deformation of the non-unitary minimal models $\mathcal M_{2,2n+3}$ --the first of which is the Lee-Yang model-- and investigate its simplest possible CDD deformation,
\beq
    \Phi(\theta) = \frac{\ii\sin u + \sinh\theta}{\ii\sin u - \sinh\theta}\,,
    \label{eq:bosonic_limit}
\eeq
in the limit $u\rightarrow 0^{-}$. In other words, recalling \eqref{eq:massless_resonance} we will effectively focus our attention on the bosonic counterparts of the $\Phi_{1,3}$ integrable deformations. 
Most curiously, at least to our knowledge these models have not been constructed previously even though they give a standard (single branch) TBA.

These $S$-matrices provide a paradigmatic example of how the iterative method to solve TBA equations can give rise to non-physical instabilities that can be circumvented applying the pseudo-arclength continuation method described in section~\ref{sec:numerical_methods}. Furthermore they offer a very different testing ground for the condition on the $L_1$ norm of the kernel \eqref{L1_norm} being the determining factor for the appearance of a critical radius. 

Let us now briefly review some properties of the models. The integrable theories resulting from the $\Phi_{1,3}$ deformation have a spectrum of $n$ particles of mass $m_a=\sin\(\frac{a\pi}{2n+1}\)$, $a=1\ldots n$. The scattering amplitudes defining the interactions between them can be obtained with the usual integrable bootstrap and --contrary to the $S$-matrices in the previous section-- have bound-state poles. The scattering matrices read
\beqa 
S_{11}(\theta) &=& \text{th}_{\frac{2}{2n+1}} (\theta)\,,\\
S_{ab}(\theta) &=& \text{th}_{\frac{|a-b|}{2n+1}} (\theta)\, \text{th}_{\frac{a+b}{2n+1}} (\theta) \prod\limits_{k=1}^{\text{min} (a,b)-1} \[\text{th}_{\frac{|a-b|+2k}{2n+1}} (\theta)\]^2\,,\la{S_minmodels}
\eeqa
where the building blocks $\text{th}_x(\theta)$ are simple CDD poles $\text{th}_x(\theta)= \frac{\tanh\[(\theta+i\pi x)/2\]}{\tanh\[(\theta-i\pi x)/2\]}=\frac{\sinh(\theta)+i \sin(\pi x)}{\sinh(\theta)-i \sin(\pi x)}$, see  \cite{Mussardo:2020rxh} and references therein.

Following the logic of the previous section, we will study these fermionic models and their bosonic counterparts defined purely from their $S$-matrix $S_{ab}^\text{b}(\theta)=-S_{ab}(\theta)$. The TBA equations are given by
\beq\label{tba_minimal}
\epsilon_a(\theta) = m_a r \cosh\theta \mp \sum\limits_{b=1}^n\, \int\limits_{-\infty}^\infty \varphi_{ab}(\theta-\theta') \ln\[1 \pm e^{-\epsilon_b(\theta')}\] \frac{d\theta'}{2\pi} \qquad\text{for}\qquad a = 1, \dots, n\,,
\eeq
and while this is now a coupled system of $n$ equations over the unknown pseudo-energies $\epsilon_a(\theta)$ for $a = 1, \dots, n$, all of the numerical techniques described in section~\ref{sec:numerical_methods} can be straightforwardly generalized to account for this change.

If we solve the TBA equations \eqref{tba_minimal} iteratively, we find that for $n>2$ (or $n>1$ in the bosonic case) there is an instability at a finite radius $r_* > 0$. For all values of $n$ we notice that the bosonic case displays this instability before the fermionic one. We could suspect that this indicates a true physical critical radius, but switching to the pseudo-arclength continuation method proves this to be false. Indeed, there is nothing strange about these models and we can continue the solutions all the way to $r=0$, reproducing the well-known effective central charge in the UV for the fermionic models.

\begin{figure}[ht!]
     \centering
     \begin{subfigure}[b]{0.75\textwidth}
         \centering
         \includegraphics[width=\textwidth]{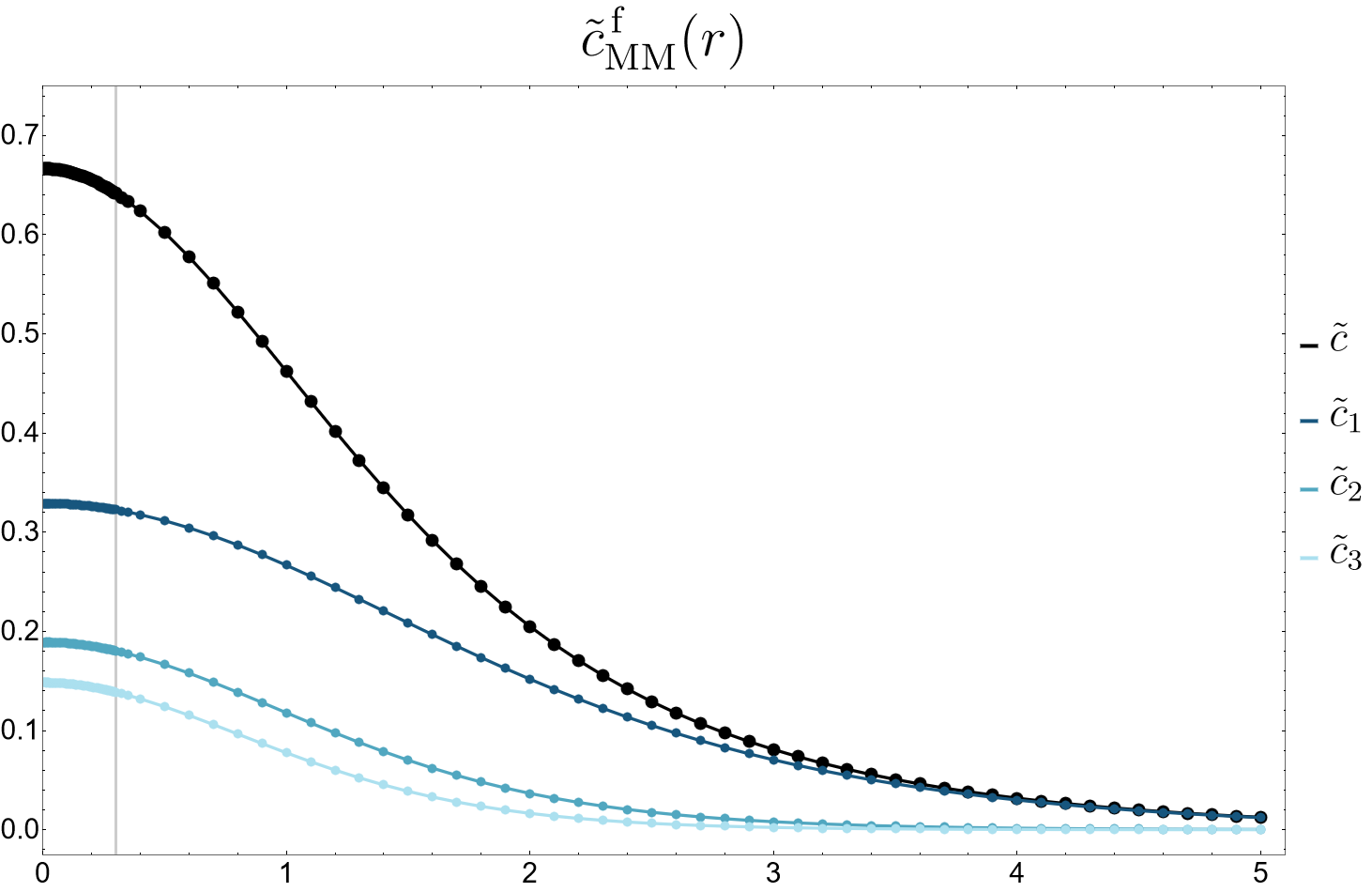}
     \end{subfigure}
     \hfill
     \begin{subfigure}[b]{0.75\textwidth}
         \centering
         \includegraphics[width=\textwidth]{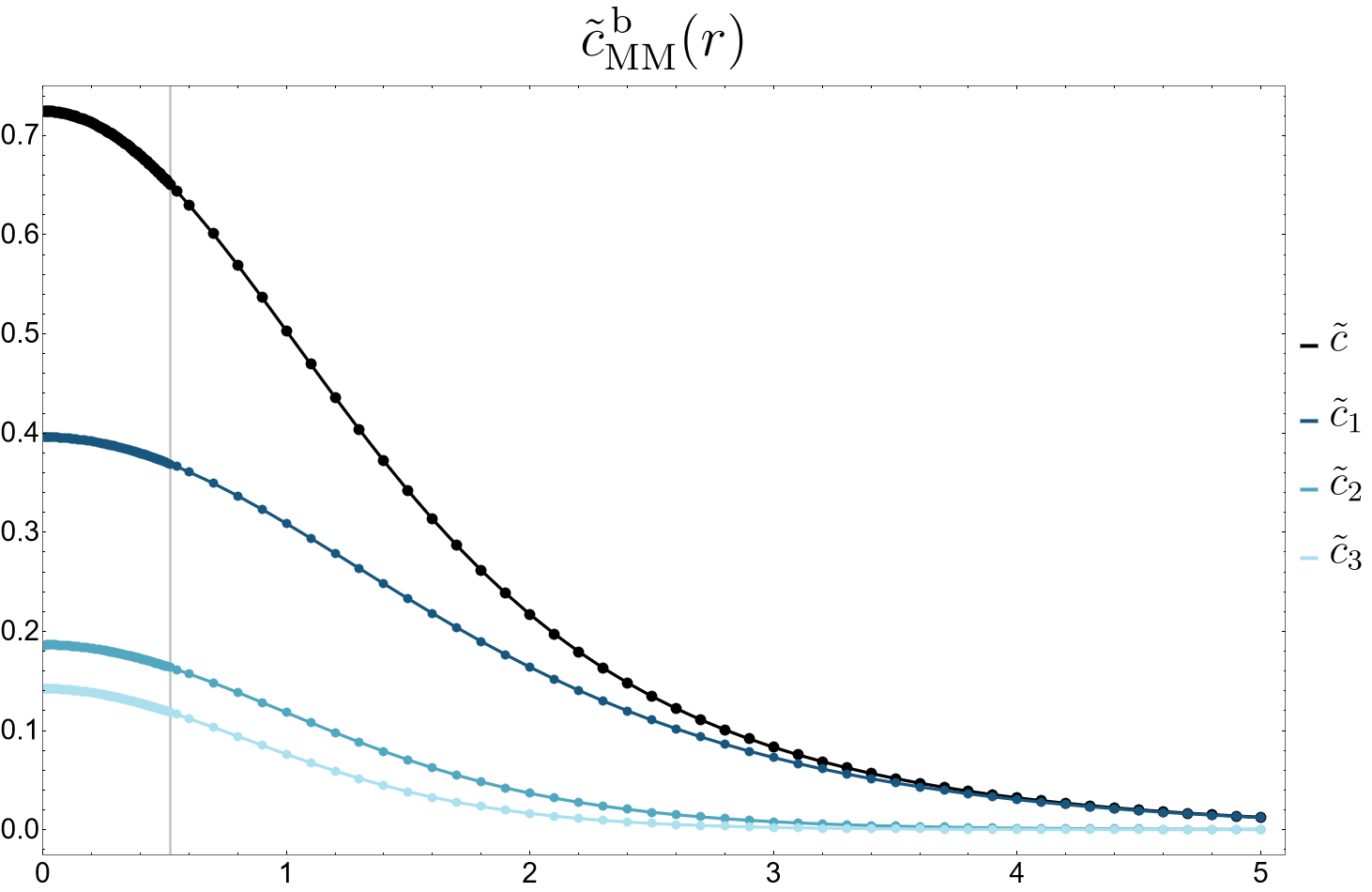}
     \end{subfigure}
        \caption{Effective central charge (black) for the  $\Phi_{1,3}$ deformed minimal model $\mathcal M_{2,2n+3}$ with $n=3$, in its fermionic (top) and bosonic (bottom) flavours. The blue curves refer to the single particle contributions $\tilde c_a(r)$ as in equation \eqref{c_a}. The vertical gray line shows the radius $r_*$ where the iterative method becomes unstable. For $r<r_*$ we use the pseudo arclength numerical method. The effective central charges converge in the ultraviolet to $2/3$ in the fermionic model and to $0.724253$ in the bosonic one.}
        \la{fig_min_models}
\end{figure}

Looking closer, it turns out that below $r_*$ in the iterative method even and odd iterations converge to different values.\footnote{Note that none of these values are true solutions to the TBA equations, since we have $$\epsilon_a^\text{even,odd}(\theta) = m_a r \cosh\theta \mp \sum\limits_{b=1}^n\, \int\limits_{-\infty}^\infty \varphi_{ab}(\theta-\theta') \ln\[1 \pm e^{-\epsilon_b^\text{odd,even}(\theta')}\] \frac{d\theta'}{2\pi}\,.$$} This is a purely numerical artifact which we can expect to avoid using a variation of the iterative algorithm with weighted-updates, \textit{i.e.} replacing \eqref{iterative_update} with
\beq
\epsilon^{(k+1)}_a(\theta) = (1-\alpha)\epsilon_a^{(k)}(\theta) + \alpha\left(
m_a r \cosh\theta \mp \sum\limits_{b=1}^n\, \int\limits_{-\infty}^\infty \varphi_{ab}(\theta-\theta') \ln\[1 \pm e^{-\epsilon_b^{(k)}(\theta')}\] \frac{d\theta'}{2\pi}\right)\,,
\eeq
for $a=1, \dots, n$, $k=0,1,\dots$ and some $\alpha \in(0,1)$. In practice, $\alpha = 1/2$ works well, and we could skip the pseudo-arclength continuation altogether. Note however the latter is a more robust stabilization method for the TBA equation, as seen in the previous section.

Of course, the fact that there is no critical radius is also expected from the observation that in this theory there are no resonances and only bound states. Indeed, a condition on the norm of the kernel in the spirit of \eqref{L1_norm} would not be satisfied, so that we should not find more than one branch in this theory.

The effective central charge in these models receives contributions from each particle in the following way
\beq\la{c_a}
\tilde c(r)=\sum\limits_{a=1}^n \tilde c_a(r)=\sum\limits_{a=1}^n\pm\frac{6}{\pi}\,\frac{m_a}{m_1}\int\limits_{-\infty}^{\infty}  r \cosh(\theta)\ln\[1\pm e^{-\epsilon_a(\theta)}\]\frac{d\theta}{2\pi}\,.
\eeq

In figure~\ref{fig_min_models} we show the fermionic (top) and bosonic (bottom) effective central charges for $n=3$. In the fermionic TBA, the effective central charge at $r=0$ converges to the expected values for $\mathcal M_{2,2n+3}$ minimal models, given by the formula $\tilde c^{(n)}=2n/(2n+3)$. In the bosonic case the UV effective central charge is given by larger values in the range $[1/2,1)$, which for $n>1$ appear to be irrational and for which we have not found an analytic expression.\footnote{For the curious reader we report $\tilde c(r=0)\approx0.5,0.641304, 0.724253, 0.778979, 0.817083$ for $n=1,2,3,4,5$\,.} It would be very interesting to see if there exists a simple realization of these bosonic models.

\section{Discussion}\label{sec:discussion}

In this work we have analyzed the Thermodynamic Bethe Ansatz for various integrable models defined through consistent $2\rightarrow2$ scattering matrices. With the exception of the results discussed in section~\ref{sec:minimal_models}, we concentrated on theories with a single stable particle and no bound states. We studied the sinh-Gordon model, which has one resonance, its elliptic deformation, with infinite resonances, and in-between models with $k$ resonances described by $k$ CDD-zeros. We have taken a bottom-up approach and studied both the fermionic and the lesser studied bosonic realizations of the TBA equations for all of the above models.

Following \cite{Camilo:2021gro}, we have confirmed that the determining factor for the TBA to have a turning point at a critical length scale $r_c$ is the $L_1$ norm of the kernel, which measures the difference between the number of bound states and resonances in the theory. Namely, all the models with a single stable particle and any number of resonances --with the exception of the single resonance fermionic model, \textit{i.e.} sinh-Gordon-- have a bifurcation point. The exact length scale at which the bifurcation happens depends on the details of the model.

An important result in the present work is that we were able to continue the TBA solutions past the turning point and follow the pair of complex conjugate solutions all the way to the UV regime. This required an efficient implementation of the pseudo-arclength continuation method to solve the (complexified) TBA equations. We found that the complex solutions to the TBA equations are such that the UV effective central charge is minimized. Moreover, we showed that the latter decreases with the number of resonances and finally goes to zero as we consider an infinite number of them, as in the elliptic sinh-Gordon models.

The existence of a turning point in the TBA --interpreted as the presence of a finite Hagedorn temperature-- and of a complex effective central charge at short distances implies a non-Wilsonian UV behaviour for these theories, just as it is the case in the $\TTB$-deformed theories and in their generalizations \cite{Camilo:2021gro}. Our results shed light on the possible nature of the theories with infinite resonances ubiquitous in the $S$-matrix bootstrap, and add to the body of experimental evidence substantiating the expectation that the majority of self-consistent $S$-matrices are not derivable from local QFT.
Indeed, there are further hints that some of the consistent $S$-matrices found saturating bounds cannot describe conventional UV-complete theories. In particular, for the $O(N)$ periodic Yang-Baxter (describing a global symmetry generalization of our infinite resonance model) there are two concrete pieces of evidence. First, in \cite{Karateev:2019ymz} it was argued with an $S$-matrix/Form Factor bootstrap involving c-minimization that this theory seems incompatible with a standard UV fixed point. Secondly, the authors of \cite{Gorbenko:2020xya} observed walking behaviour \cite{Gorbenko:2018ncu,Gorbenko:2018dtm} in this model for $N>2$ and proposed a relation to complex CFTs.

With the above hints and the fact that we are obtaining complex effective central charges, it is very tempting to think of the underlying UV theories as complex CFTs. This is an intriguing idea that warrants further investigation. A first step would be to extract the scaling dimensions of primary operators by considering the excited state TBA equations obtained by analytically continuing the groundstate ones \cite{Dorey:1996re}. Regardless of the details of the underlying UV theories, we believe that analyzing theories outside the paradigm of a standard UV fixed point is a worthy endeavour. Moreover, while our work focused on integrable theories we expect their qualitative properties to be of a universal nature, applying equally well to non-integrable and non-Wilsonian QFTs, just as it is the case for the $\TTB$ deformation.

Finally, the results discussed in section~\ref{sec:minimal_models} constitute a first incursion into the realm of CDD deformations of theories with an arbitrary number of stable particles. Specifically we studied the bosonic counterparts of the $S$-matrices describing the $\Phi_{1,3}$ deformed non-unitary minimal models $\mathcal M_{2,2n+3}$. We found that these theories possess a standard Wilsonian UV limit, corresponding to CFTs whose central charges appear to be irrational numbers between $1/2$ and $1$. We find it remarkable that these models have never, to the extent of our knowledge, appeared before in the literature. We believe it would be interesting to look at the bosonic counterparts of other well-known theories --\textit{e.g.} the unitary minimal models-- as well as to search for their QFT construction. The next natural step in this direction is to analyze a general family of theories having $n$ bound states and $N$ resonances. According to the argument we have already put forward before, we expect these to have a standard Wilsonian UV behavior for $n\geq N$, and conversely to develop a Hagedorn temperature for $n<N$. We hope to return to this point in a future publication.

\section*{Acknowledgments}
We would like to thank Yifei He, Alexandre Homrich, Slava Rychkov, Roberto Tateo, Alexander Zamolodchikov and Bernardo Zan for useful and insightful discussions. Work of SN is supported by NSF grant PHY-1915219.

This project has received funding from the European Research Council (ERC) under the European Union's Horizon 2020 research and innovation programme (QUASIFT grant agreement 677368).
\pagebreak

\appendix


\section{Some properties of elliptic functions}\label{app_ellipt}
Elliptic functions are meromorphic functions in the complex plane with two periods $\omega_1,\omega_2$ satisfying $\text{Im}\(\omega_1/\omega_2\)>0$. In this appendix we review some of their nice properties.\footnote{There are many textbooks on the theory of elliptic functions, here we follow \cite{chandrasekharan2012elliptic}.}

\paragraph{Poles and zeros}
Any non-constant elliptic function must have at least one pole in a period parallelogram. The sum of all residues at the poles inside this parallelogram is zero. Therefore, any non-constant elliptic function has at least order two: inside the period parallelogram we have either two simple poles with opposite residues or a double pole with zero residue. The former case gives rise to Jacobi elliptic functions while the latter corresponds to Weierstrass elliptic functions. Any elliptic function can be expressed in terms of Jacobi or Weierstrass elliptic functions.

\paragraph{Algebraic relations} Two elliptic functions $X(\theta)$, $Y(\theta)$ with the same periods $\omega_1,\,\omega_2$ satisfy an algebraic relation of the form
\beq
F(X,Y)=\sum\limits_{\substack{i,j=0 \\ i+j\leq n}}^n c_{i j} X^i Y^j=0\,. \label{alg_reln}
\eeq
In the following we describe how these polynomial relations arise and in particular how the total degree of the polynomial is related to the order of the poles in $X(\theta)$, $Y(\theta)$.

Let $a_1,a_2,\ldots,a_m$ be the set of distinct poles that belong to at least one of the functions. The functions might have a pole at the same position with different orders,
\beq
X(\theta)\sim\frac{r_k}{(\theta-a_k)^{m_k}}\,,\quad Y(\theta)\sim\frac{r'_k}{(\theta-a_k)^{n_k}}\,.
\eeq
We define the total pole order $K$ as the sum of the largest order for each pole,
\beq
K=\sum\limits_{k=1}^m \text{max}(m_k,n_k)\,.
\eeq
Now let us consider an elliptic function $\Phi(\theta)$ defined by a polynomial with no constant term
\beq
\Phi(\theta)=\sum\limits_{\substack{i,j=0 \\ 0<i+j\leq n}}^n c_{i j} X^i Y^j\,.
\eeq
The idea is to fix the coefficients $c_{i,j}$ such that $\Phi(\theta)$ is an entire function and therefore equals a constant by Liouville's theorem.\footnote{This theorem holds because the function would also be bounded in the period parallelogram with vertices $\{0,\omega_1,\omega_1+\omega_2,\omega_2\}$.} The function $\Phi(\theta)$ can have poles only at $a_1,a_2,\ldots,a_m$, so for it to reduce to a constant the principal parts near these points must all vanish. For example, close to $a_k$ the function $\Phi(\theta)$ has a Laurent expansion
\beq
\Phi(\theta)=\sum\limits_{l=0}^\infty A_l (\theta-a_k)^l+\sum\limits_{l=1}^{n\times \text{max}(m_k,n_k)} B_l (\theta-a_k)^{-l}\,.
\eeq
We require that the $B_l$ coefficients in the second sum vanish for all
$l$. In this way we get $n\times \text{max}(m_k,n_k)$ equations $B_l\[c_{ij},X(a_k),Y(a_k)\]=0$. Repeating this procedure for all poles gives at most $n\sum_k\text{max}(m_k,n_k)=n K$ linear homogeneous equations with variables $c_{ij}$. For a non-trivial solution to exist we need to have more unknowns than equations, \textit{i.e.} $n(n+3)/2>nK$. This leads to a bound on the total degree of the polynomial in terms of the total pole order, $n>2K-3$.

Having determined the coefficients $c_{ij}$ we can now write $\Phi(\theta)=-c_{00}$, from which we recover a non-trivial algebraic equation as \eqref{alg_reln}.

\section{Approximate algebraic relation between kernel and convolution}\label{app:algebraic_relation}

A curious fact stems from the otherwise trivial observation that in the elliptic sinh-Gordon models the convolution term in the TBA equation \eqref{tba_equation} has the same real period as the kernel $\varphi_{a,l}(\theta)$. If these functions were to share their imaginary period as well it would imply an algebraic relation between them, as explained in appendix~\ref{app_ellipt}. Given the analytic structure of the kernel, we would expect this relation to be of at least second order, \textit{i.e.} to be of the form
\beq\label{algebraic_relation}
c_{00} + c_{10}\, \mathcal{C}(\theta) + c_{01}\,
\varphi_{a,l}(\theta) + c_{11}\, \mathcal{C}(\theta) \varphi_{a,l}(\theta) + c_{20}\, \mathcal{C}(\theta)^2 + c_{02}\, \varphi_{a,l}(\theta)^2 = 0\,,
\eeq
where $\mathcal{C}(\theta) = \epsilon(\theta) - r \cosh(\theta)$. Note that if such a relation holds, it would not only require some physical interpretation but also be of practical importance, since it would permit a much more efficient analysis and solution of the TBA equation.

\begin{figure}[h!]
\centering
\includegraphics[width=.95\textwidth]{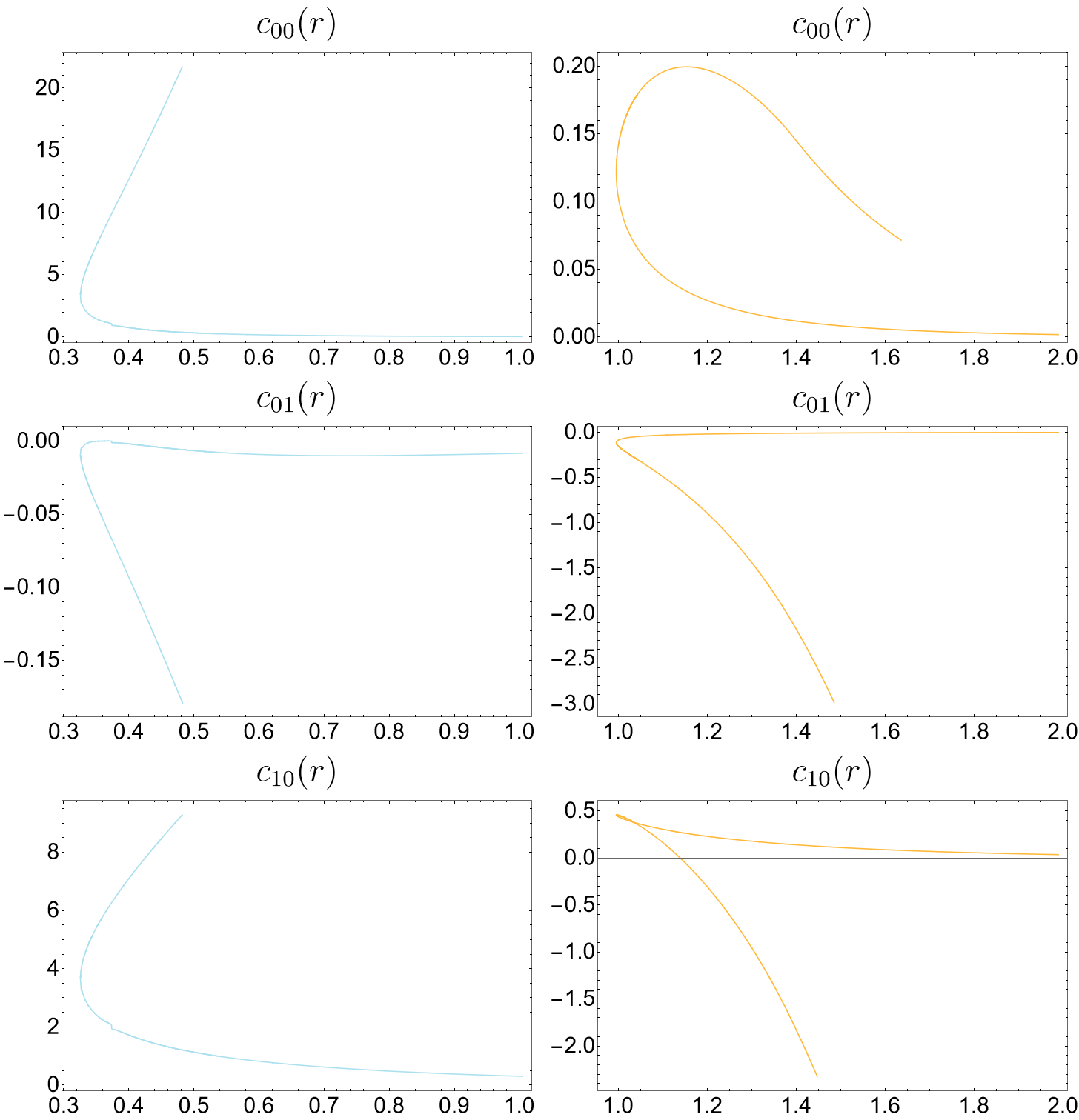}\vspace{0.3cm}
\caption{Non-vanishing coefficients in the algebraic relation \eqref{algebraic_relation} between the kernel $\varphi_{a,l}(\theta)$ and the convolution $\mathcal{C}(\theta) = \epsilon(\theta) - r \cosh \theta$ of the elliptic sinh-Gordon models with $a = 1/2$ and $l = 1/2$. Panels on the l.h.s. correspond to the fermionic model, whereas panels on the r.h.s correspond to bosonic model.}
\label{fig:algebraic}
\end{figure}

Having at our disposal numerical solutions for the convolution, determining the coefficients $c_{ij}$ reduces to performing a linear regression on the $2 N M$ datapoints \eqref{theta_discretization} to minimize the mean squared deviation to \eqref{algebraic_relation}. Restricting ourselves to the real branches, we find in practice that $c_{11} = c_{02} = 0$, so that fixing the normalization through $c_{20} = 1$ we are left with only three undetermined coefficients, $c_{00}$, $c_{10}$ and $c_{01}$. In our experiments with $N = 200$ and $M = 2$, fitting these three parameters produces residuals of relative magnitude $\mathcal{O}(10^{-5})$ for all real solutions. This is very close to the accuracy of the solutions themselves, therefore we can assert that the algebraic relation \eqref{algebraic_relation} is indeed satisfied to the numerical precision we are working with. For illustration purposes, in figure~\ref{fig:algebraic} we display plots of the non-zero coefficients as a function of $r$, for the fermionic (left panels) and bosonic (right panels) models at $a = 1/2$ and $l = 1/2$.

While we can verify the algebraic relation \eqref{algebraic_relation} holds for the real branch of solutions to the TBA equations, we were unable to do the same for the solutions belonging to the complex branch. In fact, in order to prove this relation, it would be necessary to establish the analytic structure of the convolution in the complex plane, as well as the non-trivial periodicity in the imaginary direction. This involves a non-trivial analytic continuation of the TBA solutions from the real rapidity line into the complex plane, which we are not able to do reliably all the way to $\theta\in\mathbb R+2\pi i$ as needed to confirm the validity the algebraic relation. We leave a deeper exploration of this interesting curiosity to future work.

\section{Pseudo-arclength continuation of the TBA equation}
\label{app:pseudo-arclength}

In this appendix we explain in more detail the application of the pseudo-arclength continuation method to the TBA equation \eqref{tba_equation}. After introducing the discretization \eqref{theta_discretization}, we have to solve
\beq\label{tba_equation_discretized}
\epsilon_i = r \cosh\theta_i \mp \frac{\Delta\theta}{2\pi}\sum_{j=1}^{2 N M} (\varphi_{a,l})_{ij} \ln\left(1 \pm e^{-\epsilon_j}\right) \qquad\text{for}\qquad i=1,2,\dots, 2NM\,,
\eeq
where $\Delta \theta = \Tl/N$, $(\varphi_{a,l})_{ij} = \varphi_{a,l}(\theta_i - \theta_j)$ and $\epsilon_i = \epsilon(\theta_i)$ are $2NM$ unknowns. We can accomplish this task using standard Newton-like iterative methods broadly applicable to the solution of nonlinear equations, as long as we can provide a good-enough initial starting point for the discretized pseudo-energy $\epsilon_i$.

We consider the situation where the parameters $a$ and $l$ in the kernel above are fixed beforehand, and we want to solve this problem repeatedly for various values of $r$. In this setup, small changes in $r$ and the $\epsilon_i$ are related by
\beq\label{delta_tba_equation}
\delta\epsilon_i = \delta r \cosh\theta_i + \frac{\Delta\theta}{2\pi}\sum_{j=1}^{2 N M} (\varphi_{a,l})_{ij} \frac{\delta\epsilon_j}{e^{\epsilon_j} \pm 1} \qquad\text{for}\qquad i=1,2,\dots, 2NM\,.
\eeq
If we now take $(\delta\epsilon_i, \delta r)$ to be unknown, we need to introduce an additional equation through the normalization condition
\beq\label{arclength_normalization}
\frac{1}{2NM} \sum_{i=1}^{2NM} \delta\epsilon_i^2 + \delta r^2 = 1\,,
\eeq
which leaves us with $2NM+1$ equations as required. Once again, providing a starting point for $(\delta\epsilon_i, \delta r)$ that is close enough to the actual solution is all that is necessary to solve these equations applying standard Netwon-like methods. 

Assuming we already have a sequence of $n$ solutions to \eqref{tba_equation_discretized} we wish to extend,\footnote{These may come, for example, from the application of the iterative method described in section~\ref{sec:numerical_methods}.} say $\{\epsilon_i^{(1)}, \epsilon_i^{(2)}, \dots, \epsilon_i^{(n)}\}$ corresponding to $\left\{r^{(1)}, r^{(2)}, \dots, r^{(n)}\right\}$, we can use its last two elements to provide an initial guess for $(\delta\epsilon_i, \delta r)$,
\beq
\delta\epsilon_i \approx \mathcal{N} (\epsilon_i^{(n)}-\epsilon_i^{(n-1)}) \qquad\text{and}\qquad \delta r \approx \mathcal{N} (r^{(n)}-r^{(n-1)})\,,
\eeq
where the normalization constant $\mathcal N$ is given by
\beq
\mathcal{N} = \left(\frac{1}{2NM} \sum_{i=1}^{2NM} (\epsilon_i^{(n)}-\epsilon_i^{(n-1)})^2 + (r^{(n)}-r^{(n-1)})^2\right)^{-1/2}\,.
\eeq
Solving the derivative equation \eqref{delta_tba_equation} and normalization condition \eqref{arclength_normalization} for $(\delta\epsilon_i, \delta r)$ then allows us to construct an approximate solution to the original TBA equation. Indeed, introducing a \textit{pseudo-arclength} parameter $\Delta s$ controlling how far we want to move away from the last available solution $\epsilon_i^{(n)}$, we can use $\epsilon_i \approx \epsilon_i^{(n)} + \delta \epsilon_i \Delta s$ as an initial guess for the solution of \eqref{tba_equation_discretized}. This corresponds to $r \approx r^{(n)} + \delta r \Delta s$, therefore we need to promote $r$ to become an $(2NM+1)$-th unknown, while introducing an additional constraint
\beq\label{arclength_constraint}
\frac{1}{2NM} \sum_{i=1}^{2NM} \delta\epsilon_i (\epsilon_i - \epsilon_i^{(n)}) + \delta r (r-r^{(n)}) = \Delta s\,.
\eeq
Thus, the solution of \eqref{tba_equation_discretized} and \eqref{arclength_constraint} provides a new term $\epsilon_i^{(n+1)}$ in our sequence, and the process can be repeated \textit{ad arbitrium}.

\bibliography{biblio}

\end{document}